\UseRawInputEncoding 
\documentclass[10pt,journal,compsoc]{IEEEtran}

\usepackage{booktabs}
\usepackage{graphicx}
\usepackage[table,xcdraw,x11names,dvipsnames]{xcolor}
\usepackage{dblfloatfix}
\newcommand{\todo}[1]{}
\renewcommand{\todo}[1]{{\color{red} TODO: {#1}}}
\usepackage{comment}
\usepackage{amssymb}
\usepackage{multirow}
\usepackage{lscape}
\usepackage{array, multirow, bigdelim, makecell}
\usepackage{fontawesome5}
\newcommand{\tabitem}{~~\llap{\textbullet}~~}
\usepackage[skins]{tcolorbox} 

\def\mybarhhigh#1#2{
   {\color{black}\rule{#1mm}{8pt}}  #2}
   
\def\mybar2#1#2#3#4#5{
   {\color{black}\rule{4pt}{#1mm}}
 {\color{black}\rule{4pt}{#2mm}}
  {\color{black}\rule{4pt}{#3mm}}
   {\color{black}\rule{4pt}{#4mm}}
    {\color{black}\rule{4pt}{#5mm}}
   }

\def\mybarX#1#2#3#4#5{
   {\color{black}\rule{4pt}{#1cm}}
 {\color{black}\rule{4pt}{#2cm}}
  {\color{black}\rule{4pt}{#3cm}}
   {\color{black}\rule{4pt}{#4cm}}
    {\color{black}\rule{4pt}{#5cm}}
   }

\def\mybarhhighY#1#2{
   {\color{black}\rule{#1mm}{8pt}}  #2}

\ifCLASSINFOpdf

\else

\fi

\begin{document}


\title{Emotion--Centric Requirements Change Handling in Software Engineering}

\author{Kashumi Madampe,~\IEEEmembership{Graduate Student Member,~IEEE,}
        Rashina~Hoda,~\IEEEmembership{Member,~IEEE,}
        and~John~Grundy,~\IEEEmembership{Senior Member,~IEEE}
\IEEEcompsocitemizethanks{\IEEEcompsocthanksitem K. Madampe, R. Hoda, and J. Grundy are with the HumaniSE Lab at Department of Software Systems and Cybersecurity, Faculty of Information Technology, Monash University, Wellington Road, Clayton, VIC 3800, Australia.\protect\\

E-mail: kashumi.madampe@monash.edu}%
\thanks{Manuscript received May 11, 2022; revised Month Date, 2021.}
\markboth{Submitted to IEEE Transactions on Software Engineering}%
}

\IEEEtitleabstractindextext{%
\begin{abstract}





\textbf{Background:}
Requirements Changes (RCs) -- the additions/modifications/deletions of functional/non-functional requirements in software products -- are challenging for software practitioners to handle. Handling some changes may significantly impact the emotions of the practitioners. 
\textbf{Objective:}
We wanted to know the key challenges that make RC handling difficult, how these impact the emotions of software practitioners, what influences their RC handling, and how RC handling can be made less emotionally challenging.
\textbf{Method:}
We followed a mixed-methods approach. 
We conducted two survey studies, with 40 participants and 201 participants respectively. The presentation of key quantitative data was followed by descriptive statistical analysis, and the qualitative data was analysed using Strauss--Corbinian Grounded Theory, and Socio--Technical Grounded Theory analysis techniques.
\textbf{Findings:}
We found (1) several key factors that make RC handling an emotional challenge, (2) varying emotions that practitioners feel when it is challenging to handle RCs, (3) how stakeholders, including practitioners themselves, peers, managers and customers, influence the RC handling and how practitioners feel due to the stakeholder influence, and (4)  practices that can be used to better handle RCs.
\textbf{Conclusion:}
Some challenges are technical and some are social which also belong to aspects of agile practice, emotional intelligence, and cognitive intelligence. Therefore, to better handle RCs with positive emotions in socio--technical environments, agility, emotional intelligence, and cognitive intelligence need to cooperate with each other.  

\end{abstract}

\begin{IEEEkeywords}
emotions, emotional intelligence, affects, requirements, changes, human factors, mixed-methods, software engineering, software teams, socio-technical grounded theory, agile, job-related affective well-being scale, well-being, workplace awareness
\end{IEEEkeywords}}

     \maketitle

\IEEEdisplaynontitleabstractindextext

\IEEEpeerreviewmaketitle

\ifCLASSOPTIONcompsoc
\IEEEraisesectionheading{\section{Introduction}\label{sec:introduction}}
\else
\section{Introduction}
\label{sec:introduction}
\fi

\IEEEPARstart{E}{motions} are a fundamental part of being human. Scherer defines emotions as \emph{``a sequence of interrelated, synchronised changes in the states of all the five organismic subsystems (information processing, support, executive, action, and monitoring) in response to the evaluation of an external or internal stimulus event as relevant to central concerns of the organism''} \cite{Scherer1987TowardEmotion}. While emotions play a vital role for human beings, in general requirements changes (RCs) play an important role in software development teams.

RCs can be additions/modifications/deletions of functional/non-functional requirements of any software \cite{Madampe2021AContexts}. As RCs are unavoidable and impact the scope, cost, and time of the software development project, RCs can often be challenging for software practitioners to handle. RCs also act as stimuli in triggering the emotions of software practitioners who handle them. Existing research recognises the association between emotions and behavior \cite{Colomo-Palacios2010AEngineering}, cognition \cite{Colomo-Palacios2010AEngineering}, productivity \cite{Colomo-Palacios2010AEngineering}, \cite{Graziotin2015DoEngineering}, \cite{Kolakowska2013EmotionEngineering}, \cite{Girardi2021EmotionsWorkplace}, and decision-making \cite{Muller2015}. 

Over the past decade, many studies have been conducted to explore the impact of emotions of software practitioners in general, and emotions and their impact on productivity \cite{Girardi2021EmotionsWorkplace}, \cite{Wrobel2016TowardsTeams}, \cite{Wrobel2013}, \cite{Crawford2014}, \cite{Graziotin2014}, \cite{Graziotin2014SoftwarePerformance}. Furthermore, in light of the recent global pandemic, there has been renewed interest in studying emotions in software engineering (SE) contexts \cite{Ralph2020PandemicProgramming}. However, less attention has been given to investigating the emotions of practitioners while handling requirements or RCs \cite{Madampe2020TowardsTeams}, \cite{Madampe2022TheEngineering}. 


We conducted two studies on RCs and emotional responses to RCs as worldwide survey studies\footnote{Approved by Monash Human Research Ethics Committee. Approval Number: 23578}, one focusing solely on RCs (survey Alpha: 40 participants) and the other on emotional responses to RCs (survey Beta: 201 participants).
Through the analysis of survey Alpha data using descriptive statistical analysis and Strauss-Corbinian Grounded Theory analysis techniques \cite{Strauss1998BasicsTechniques.}, we found several key  factors that make RC handling challenging. In survey Beta, using  Job related Affective Well-being Scale (JAWS) \cite{VanKatwyk2000UsingStressors.}, and Socio--Technical Grounded Theory for data analysis (STGT4DA) \cite{Hoda2021Socio-TechnicalEngineeringb}, we explored how practitioners feel when they find it challenging to handle RCs. We also investigated how stakeholders influence the overall handling of a RC throughout its life cycle. By synthesising findings from both survey Alpha and Beta we found the most common RC handling techniques practitioners use upon receiving RCs to prior development, i.e., pre--development RC handling techniques.
These findings ultimately led to derive a set of recommended practices that can be utilised throughout the RC handling life cycle. \\



The key contributions of this paper are:
\begin{itemize}



\item Key factors that make RC handling a challenge;
\item How practitioners feel due to RC challenges
\item How  stakeholders influence practitioners' emotions when handling RCs;
\item A number of recommendations for practitioners for a better RC handling experience; and
\item A number of future research directions for researchers.

\end{itemize}

\section{Motivation, Research Questions and Definitions}

\subsection{Motivation}

\subsubsection{Motivation from Experience}
Imagine Kash, a software developer, who unexpectedly receives an RC to work on. The RC’s high \textit{complexity}, high \textit{cascading impact}, large \textit{size}, imprecise/ unclear \textit{definition}, high \textit{priority}, high required \textit{effort}, and difficult/ irregular \textit{access to the customer} made it challenging for Kash to handle the RC. On top of this, \textit{cross--functionality} was forced within her team. All these factors make Kash feel high anxiety and low pleasurable \textit{emotions}. She thinks that if she had some pre--development techniques to use early and some best practices, these could make  handling the RC easier. This could also arouse higher pleasurable emotions in her RC handling work. We wanted to help software practitioners like Kash by better understanding their emotional responses to RCs, and then providing such guidance when they go about handling RCs. 

\subsubsection{Motivation from Related Work}
\textbf{RC Handling in Agile Contexts.} Given that agile being widely used in software development, secondary studies on agile requirements engineering, practices, and challenges highlight that studies on functional and non-functional requirements, process support and management, process quality and improvement, requirement negotiation, and acceptance tests have been studied extensively \cite{Curcio2018}. Also, studies on benefits of agile RE over traditional RE as a research area is saturated \cite{Heikkila2015}, studies on user and stakeholder involvement are in a high volume \cite{Schon2017}, and requirement prioritisation and testing before coding have been studied widely \cite{Inayat2015}. However, studies on changes in requirements \cite{Curcio2018}, and change management lacks attention in research \cite{Heikkila2015}, \cite{Schon2017}, \cite{Inayat2015}.

\textbf{Emotions in Software Engineering.} According to Muller and Fritz, the reasons for increase in emotions/progress during software development are, localising relevant code, better understanding of parts of the code, clear next steps, writing code, and having new ideas. And, the reasons for decreases in emotions/progress are, difficulty in understanding how parts of the code/API work, difficulty in localising relevant code, not being sure about next steps, realising that hypothesis on how code works is wrong, and missing/insufficient documentation \cite{Muller2015}.
Furthermore, a negative link exists between hurry and number of commits, a negative relationship exists between social interaction and hindered work well-being \cite{Kuutila2018UsingWell-being}, and emotions of a technical question impacts the probability of obtaining satisfying answers \cite{Novielli2014TowardsOverflow}. Security related discussions on GitHub contain more negative emotions than other discussions \cite{Pletea2014SecurityGitHub}. Also, frustration is felt most often during software development \cite{Wrobel2013}. This lowers productivity, while anger increases productivity, enthusiasm increases productivity, and emotions transit from frustration $\rightarrow$ anger $\rightarrow$ contentment $\rightarrow$ enthusiasm \cite{Wrobel2013}. Anxiety and nervousness are felt when presenting and satisfaction and enjoyment are felt when coding \cite{Colomo-Palacios2019EmotionsCoding}. However, emotional awareness increases developer's progress by mitigating negative emotions \cite{Fountaine2017EmotionalMeasurement}. In summary, research has been done on exploring emotions in SE in general during/ post development \cite{Murgia2018AnSystems, Yang2017AnalyzingProjects, Murgia2014DoArtifacts, Kolakowska2013EmotionEngineering, Fountaine2017EmotionalMeasurement, Novielli2018AOverflow, Neupane2019EmoD:Development, Werder2018MEME:Github, Pletea2014SecurityGitHub, Graziotin2014SoftwarePerformance, Novielli2014TowardsOverflow,Guzman2013}, including how emotions impact productivity \cite{Wrobel2013, Crawford2014, Wrobel2016TowardsTeams}, progress \cite{Girardi2018SensingExperiment, Muller2015}, and how practices impact the emotions \cite{Colomo-Palacios2019EmotionsCoding}, relationship between emotions and problem solving \cite{Graziotin2014}, and relationship between affective states and software metrics \cite{Kuutila2018UsingWell-being}. However, research on emotions during requirements engineering activities, including while handling RCs is extremely limited \cite{Colomo-Palacios2011UsingEngineering, Madampe2020TowardsTeams, Madampe2022TheEngineering}.

\subsection{Research Questions}
The key research questions that we wanted to answer in this study are:

\begin{itemize}
    \item[\textbf{RQ1.}] \textbf{What are the factors that make RC handling a challenge?}
    We were interested in understanding ``what'' makes the RC handling a challenge. We conducted survey Alpha to answer this research question.
    
    \item[\textbf{RQ2.}]\textbf{How do practitioners feel when it is challenging to handle RCs?}
    We wanted to know how practitioners feel various emotions when they handle RCs. The open--ended questions in Survey Beta resulted in details about how practitioners feel when it is challenging for them to handle their RCs.
    
    \item[\textbf{RQ3.}]\textbf{How do stakeholders influence the handling of RCs emotionally and how do software practitioners feel about this influence?}
    The open--ended questions in survey Beta illuminated how stakeholders influence the overall handling of RCs, and how practitioners feel about their influence on handling of RCs.  As we identified a range of key stakeholders -- practitioner, peers in the team, manager, and customers -- we further breakdown this research question to sub--research questions:
    
    \begin{itemize}
    \item[\textit{RQ3.1.}]\textit{How do practitioners themselves influence the handling of RCs?}
    \item[\textit{RQ3.2.}]\textit{How do peers in the team influence the handling of RCs?}
    \item[\textit{RQ3.3.}]\textit{How do team managers influence the handling of RCs?}
    \item[\textit{RQ3.4.}]\textit{How do customers influence the handling of RCs?}
    \end{itemize}
    
    \item[\textbf{RQ4.}]\textbf{How do practitioners approach handling RCs while managing emotions?}
    We asked about the techniques practitioners use to handle RCs and make their RC handling more emotionally easier. This emerged from the answers given to the open--ended questions in both surveys.
    
\end{itemize}

\subsection{Definitions}
We use the terms presented in Table \ref{tab:definitions} 
throughout the paper. The cited definitions are directly from the sources and not paraphrased.

\begin{table}[]
\caption{Definitions of Key Terms Used}
\label{tab:definitions}
\resizebox{\columnwidth}{!}{%
\begin{tabular}{@{}ll@{}}
\toprule
\textbf{Term}                   & \textbf{Definition}                                                                                                                                                                                                                                                                                                                                               \\ \midrule
Agility                & \begin{tabular}[c]{@{}l@{}}Carrying out agile practices such as collaboration, \\ self--organisation, and cross--functionality\end{tabular}                                                                                                                                                                                                              \\ \midrule
Cognitive intelligence & \begin{tabular}[c]{@{}l@{}}One’s abilities to learn, remember, reason, solve \\ problems, and make sound judgments, particularly \\ as contrasted with emotional intelligence \cite{VandenBos2007APAPsychology}\end{tabular}                                                                                                                                                               \\ \midrule
Emotion                & \begin{tabular}[c]{@{}l@{}}A sequence of interrelated, synchronised changes \\ in the states of all the five organismic subsystems \\ (information processing, support, executive, action, \\ and monitoring) in response to the evaluation of an \\ external or internal stimulus event as relevant to \\ central concerns of the organism \cite{Scherer1987TowardEmotion}\end{tabular} \\ \midrule
Emotional intelligence & \begin{tabular}[c]{@{}l@{}}Type of intelligence that involves the ability to \\ process emotional information and use it in\\ reasoning and other cognitive activities \cite{VandenBos2007APAPsychology}\end{tabular}                                                                                                                                                                      \\ \midrule
Emotion regulation     & \begin{tabular}[c]{@{}l@{}}Any process that decreases, maintains, or \\ increases emotional intensity over time, thereby \\ modifying the spontaneous flow of emotions \\ \cite{Koval2015EmotionInertia}, \cite{Gross2007EmotionPress.}, \cite{Koole2009TheReview}\end{tabular}                                                                                                                                                                     \\ \midrule
Emotion response       & \begin{tabular}[c]{@{}l@{}}An emotional reaction, such as happiness, fear, \\ or sadness, to give a stimulus \cite{VandenBos2007APAPsychology}\end{tabular}                                                                                                                                                                                                                                \\ \midrule
Empathy                & \begin{tabular}[c]{@{}l@{}}Understanding a person from his or her frame of \\ reference rather than ones own, or vicariously \\ experiencing that persons feelings, perceptions,\\ and thoughts \cite{VandenBos2007APAPsychology}\end{tabular}                                                                                                                                             \\ \midrule
Requirements Change    & \begin{tabular}[c]{@{}l@{}}Additions/modifications/deletions of functional/\\ non-functional requirements in a software project \cite{Madampe2021AContexts}\end{tabular}                                                                                                                                                                                                             \\ \bottomrule
\end{tabular}%
}
\end{table}


\section{Study Design}

\begin{figure*}
    \centering
    \includegraphics[width=\textwidth]{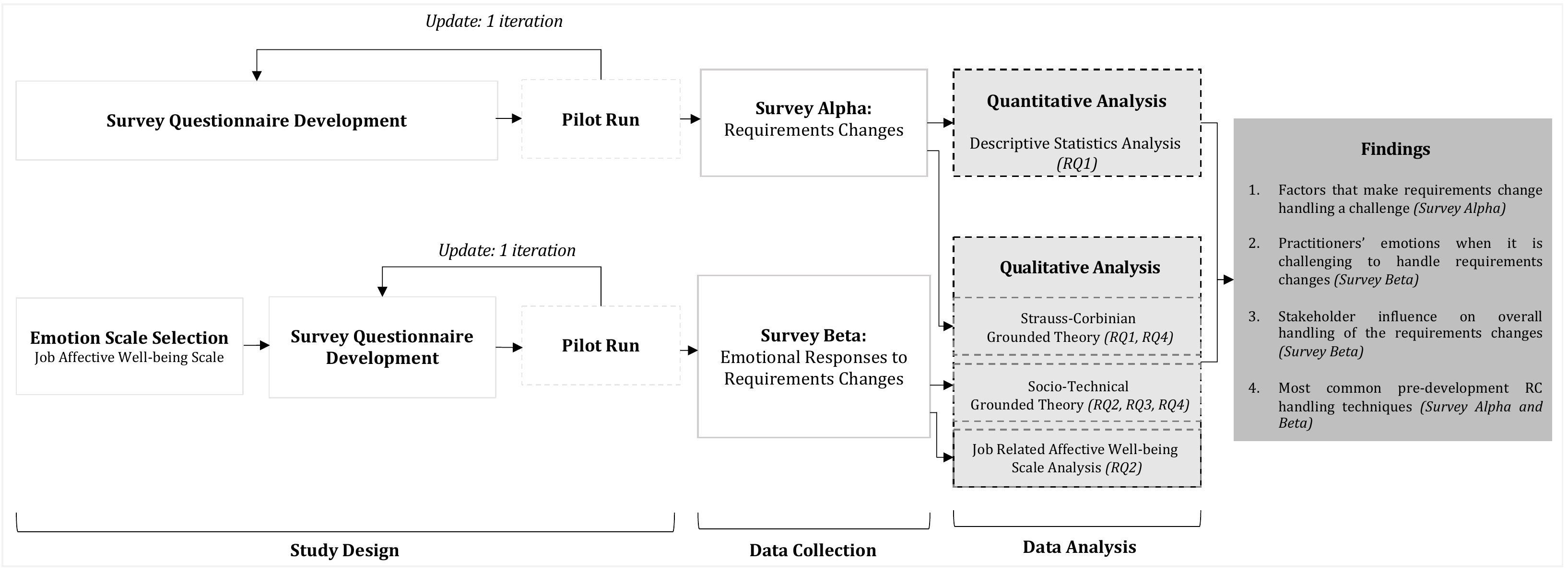}
    \caption{Our Approach}
    \label{fig:method}
\end{figure*}

We conducted two studies (survey Alpha and Beta) to gain an in--depth understanding of how RCs arise during software development (Alpha) and how practitioners emotionally respond to these RCs (Beta). 
Replication packages including the questionnaires and data for both survey Alpha and survey Beta are available online\footnote{https://github.com/users/kashumi-m/projects/1}. Fig \ref{fig:method} outlines the conduct of the studies, further explained in detail below and in the upcoming sub--sections. 


\textbf{Survey Alpha.} First, we developed the survey questionnaire (Section \ref{sec:alpha_mtd}), then conducted a pilot study to receive feedback and better refine the survey questions. Then we collected data on \textit{RCs} (Section \ref{sec:data_collection}). The collected data followed a descriptive statistical analysis (quantitative data) and Strauss--Corbinian Grounded Theory analysis (qualitative data) (Section \ref{sec:data_analysis}). The findings from this analysis answered RQ1.

\textbf{Survey Beta.} Survey Beta started with selecting an emotion scale appropriate for our study (Section \ref{sec:beta_mtd}). Then, similar to survey Alpha, we developed the survey (Section \ref{sec:beta_mtd}), and conducted the pilot study to refine the survey questionnaire. After this step, we collected data on \textit{emotional responses to RCs} (Section \ref{sec:data_collection}). We then used descriptive statistical analysis (quantitative data), and Socio--Technical Grounded Theory analysis and Job related Affective Well--being Scale (qualitative data) to analyse the data. This analysis resulted in answering RQ2 and RQ3.







\subsection{Survey Questionnaire Development}
We developed both questionnaires of survey Alpha and Beta by following Kitchenham et al.'s \cite{Kitchenham2008PersonalSurveys}, \cite{Kitchenham2002PreliminaryEngineering}, and Punter et al.'s \cite{Punter2003ConductingEngineering} guidelines. Both questionnaires had questions on demographic information, project information, and team information of the participants. The rest of the questions of survey Alpha focused on RCs and the survey Beta focused on emotional responses to RCs. 
We used \textit{Qualtrics\footnote{https://www.qualtrics.com/}} as the survey platform in both cases.

\subsection{Survey Alpha} \label{sec:alpha_mtd}
\begin{figure}
    \centering
    \includegraphics[width=\columnwidth]{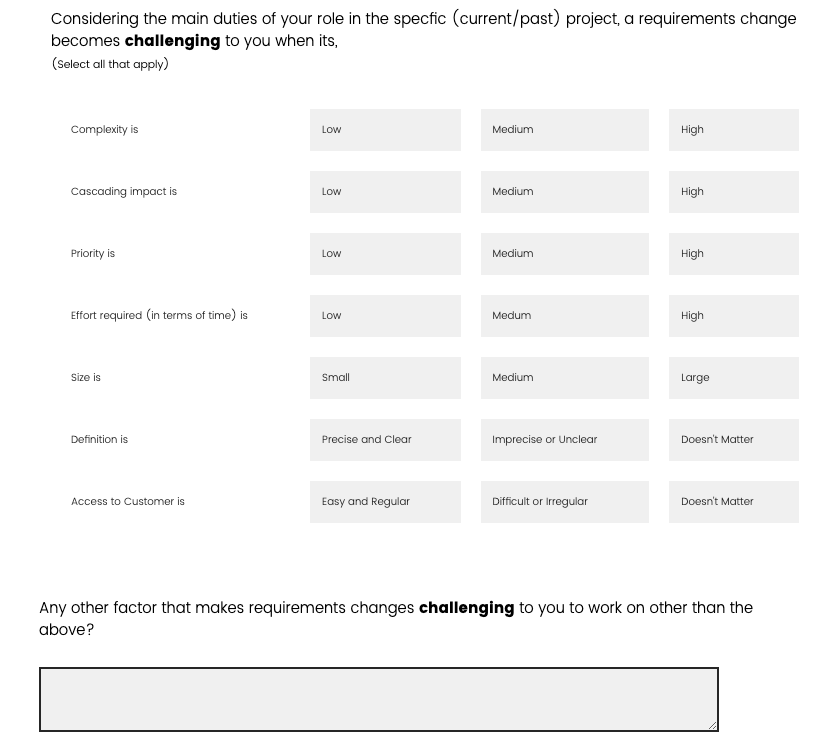}
    \caption{Survey Alpha: Question on RC Challenging Factors}
    \label{fig:alpha_challenging}
\end{figure}

The complete survey questionnaire of survey Alpha is available in the replication package. In Fig. \ref{fig:alpha_challenging} we present the question that we used to answer RQ1 and RQ4.
The participants were allowed to choose their choices from the closed--ended question, and also they had the opportunity to give their opinions through the open--ended question. 

To develop the choices for the closed--ended question, we consulted literature, our previous interview--based study, and our own collective industry experience. We considered \emph{complexity, cascading impact, size of RC, effort required, definition, priority,} and \emph{access to customer} as the factors that make RC handling challenging. According to Boehm \cite{Boehm1984SoftwareEconomics}, \textit{complexity} is one of the important drivers in software cost. We combined requirements dependability and change conflicts with existing requirements, which are considered as challenges in RC management in general \cite{Anwer2019AResults} as \textit{cascading impact}. Furthermore, we derived \emph{access to customer} from a previous interview--based study we conducted and also adapted from Hoda et al.'s work \cite{Hoda2011TheTeams} and Anwer et al.'s work \cite{Anwer2019AResults} along with \emph{prioritisation} as prioritisation is a challenge in RC management in general. Other metrics: size of RC, effort required, and definition were hypothesised based on experience. Complexity, cascading impact, effort required, and priority followed the dimensions ``low, medium, high''.

We used ``small, medium'', and ``large'' as the dimensions for \textit{size of RC}. The dimensions ``imprecise or unclear, doesn't matter'', and ``precise and clear'' were used for the factor \textit{definition}. Difficult or irregular, doesn't matter, and easy and regular were used for the factor \textit{access to customer}.

\subsection{Survey Beta} \label{sec:beta_mtd}

In order to describe and capture emotional responses to RCs, we needed a set of emotions and scale to capture them. We evaluated 20 well-established emotion scales from psychology -- the 15 as described in \cite{Curumsing2017Emotion-orientedEngineering}, as well as PANAS \cite{Watson1988DevelopmentScales.}, SPANE \cite{Diener2010NewFeelings}, JES \cite{Fisher1997}, DEQ \cite{UniversityofYork.NHSCentreforReviewsDissemination2009SystematicCare}, and JAWS \cite{VanKatwyk2000UsingStressors.}. We compared their categorisation of emotions and their applicability for use to describe practitioners' emotional responses to RCs. Through our analysis, we found 3 scales -- Discrete Emotions Questionnaire (DEQ), Job Emotion Scale (JES), and Job-related Affective Well-being Scale (JAWS) -- to be appropriate scales for use in our study. 
We found these scales as appropriate for study as the majority of the emotions given in these scales are commonly felt in software engineering contexts given our own industrial experience, whereas other scales consist of less common emotions felt in software engineering contexts. 
From our own industrial experience, we decided not to use DEQ as we came across certain emotions that were irrelevant for software development teams (e.g.: ``terror" and ``craving"). In our previous work \cite{Madampe2020TowardsTeams}, we used JES which consists of 16 emotions. However, we wanted to gain a more comprehensive understanding of emotional responses to RCs. Therefore, finally we decided to use JAWS which has been used widely to assess emotional reactions of people to their jobs. As our survey Beta questionnaire asked the participants to respond to the questions by thinking of the current or most recent project they worked on, we found JAWS likely to be the most appropriate emotion scale for our study.

JAWS has two forms: long form (30 emotions) and short form (20 emotions). We utilised the short form which the authors of JAWS claim as the form that is most commonly used \cite{Job-relatedSpector}. The 20 emotions in JAWS are categorised into 4 sub-scales along the dimensions: pleasure and arousal (intensity). The sub-scales are namely, High pleasurable-High arousal (High\textsuperscript{2}), High pleasurable-Low Arousal (High\textsuperscript{1}), Low pleasurable-High Arousal (Low\textsuperscript{1}), and Low pleasurable-Low Arousal (Low\textsuperscript{2}). We abbreviated the sub-scales as above by making the abbreviation central to the pleasure. i.e., for example, when both pleasure and arousal are high, we abbreviated it as high\textsuperscript{2}; otherwise high\textsuperscript{1}. The emotions under each sub-scale are given in Table \ref{tab:JAWS}. The scale allows the participants to select one of the following five choices choice per emotion: \textit{never, rarely, sometimes, quite often,} and \textit{extremely often}. 

\begin{table}[]
\caption{Job-related Affective Well-being Scale Sub-Scales}
\label{tab:JAWS}
\resizebox{\columnwidth}{!}{%
\begin{tabular}{ll}
\toprule
\multicolumn{1}{c}{\textbf{Sub Scale}} & \multicolumn{1}{c}{\textbf{Emotion}}                 \\ \midrule
High\textsuperscript{2}                                   & Energetic, Excited, Ecstatic, Enthusiastic, Inspired \\
High\textsuperscript{1}                                   & At-ease, Calm, Content, Satisfied, Relaxed           \\
Low\textsuperscript{1}                                  & Angry, Anxious, Disgusted, Frightened, Furious       \\
Low\textsuperscript{2}                                   & Bored, Depressed, Discouraged, Gloomy, Fatigued      \\ \bottomrule
\end{tabular}%
}
\end{table}

\begin{figure}
    \centering
    \includegraphics[width=\columnwidth]{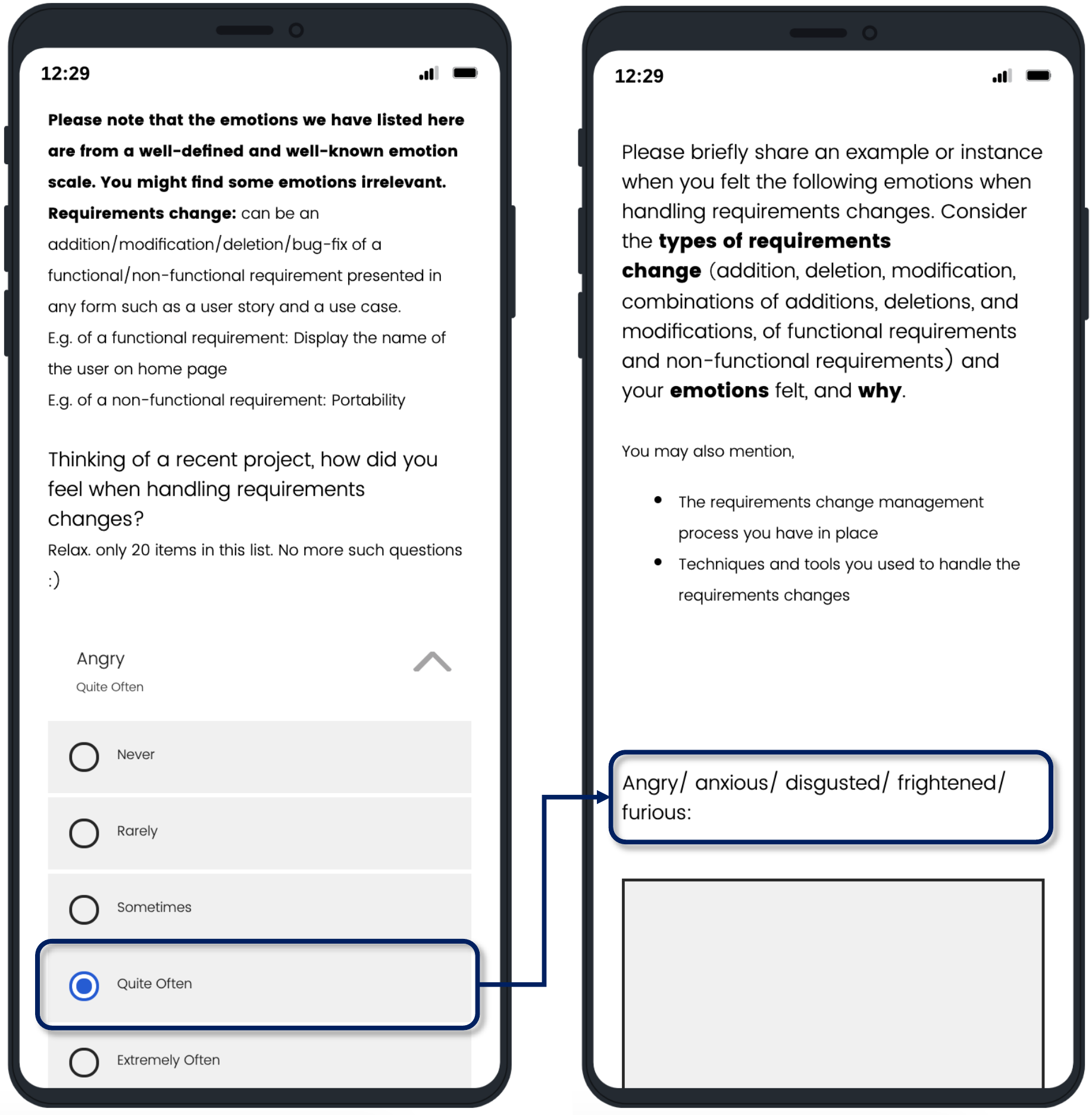}
    \caption{Survey Beta: Prompting Open--ended Questions based on the Emotions Selected in the Closed--ended Question}
    \label{fig:beta_survey}
\end{figure}

After we selected the emotion scale to use, we developed the questionnaire for survey Beta, available in the replication package. The open--ended questions that were used to answer RQ2 
followed this approach. First, we allowed participants to indicate how they feel when handling RCs through a closed--ended question (complete JAWS scale). Then, upon the selection of their emotions in that question, they were prompted with open--ended questions representing the sub--scales of the emotions in JAWS. This is illustrated in Fig. \ref{fig:beta_survey}.

\subsection{Data Collection} \label{sec:data_collection}
The data collection steps we followed during the two survey studies are summarised in Table \ref{tab:data_collection}. First we conducted pilot studies (as a part of the study design), then refined the questionnaire based on the feedback we received from pilot study participants, and finally recruited participants through various techniques. The key difference between the data collection techniques used in the surveys was the sampling methods used (DC 3.1. and DC 3.2.). In survey Alpha (40 participants), we used convenience sampling due to convenient access to participants, time and budget constraints. In survey Beta (201 participants), we used random sampling first to get a sample to represent the survey development population (as in \cite{SlashDataLtd.2019The2019} 18.9 M in 2019; sample size required=385 participants) and then purposive sampling to have fair geographical distribution and gender distribution. However, in survey Beta, we did not collect data from 385 participants as the data we collected met our requirement on saturation of quantitative data collected, i.e., when we identified the most common emotions felt during RC handling (results are given in \cite{Madampe2022TheEngineering}), where no more data collection was necessary.  We did not collect any identifiable information such as personal information from the participants, except from the participants who provided their details voluntarily for the participation in future studies.

\begin{table*}[]
\caption{Data Collection}
\label{tab:data_collection}
\resizebox{\textwidth}{!}{%
\begin{tabular}{@{}llllll@{}}
\toprule
\textbf{} & \textbf{Steps}                                                                                                                                                                                & \textbf{Survey Alpha}                                                                                                                         & \textbf{Alpha Participants}                                                                            & \textbf{Survey Beta}                                                                                                                                                                                                                   & \textbf{Beta Participants}                                                                                                                                  \\ \midrule
DC1.      & \begin{tabular}[c]{@{}l@{}}Conduct pilot study\\ \\ Qualification criteria: Experience in \\ software development industry\end{tabular}                                                       &                    \faCheckSquare                                                                                                                       & \begin{tabular}[c]{@{}l@{}}4 participants \\ (2 Research Fellows;\\ 2 PhD students)\end{tabular}   &                                                                                                                \faCheckSquare                                                                                                                     & \begin{tabular}[c]{@{}l@{}}2 participants \\ (2 PhD students)\end{tabular}                                                                               \\ \midrule
DC2.      & \begin{tabular}[c]{@{}l@{}}Refine the survey questionnaire based \\ on feedback received from \\ pilot study participants\end{tabular}                                                        & \begin{tabular}[c]{@{}l@{}}\tabitem Changed the survey title to stay \\ in layman terms\\ \tabitem Changed the estimated completion \\ time\end{tabular}    &                                                                                                    & \begin{tabular}[c]{@{}l@{}}Changed the estimated \\ completion time\end{tabular}                                                                                                                                                    &                                                                                                                                                          \\ \midrule
DC3.      & \multicolumn{5}{l}{Recruit participants}                                                                                                                                                                                                                                                                                                                                                                                                                                                                                                                                                                                                                                                                                                                                                                                                        \\ \midrule
DC3.1.    & \begin{tabular}[c]{@{}l@{}}Post the survey link on professional \\ software development groups \\ and in our profiles on social media \\ such as \textit{LinkedIn, Twitter,} and \textit{Facebook}\end{tabular} &           \faCheckSquare                                                                                                                                & \hspace{-1em}\rdelim\}{15.5}{*}[{\begin{tabular}[c]{@{}l@{}}40 participants \\ (Convenience sampling)\end{tabular}}] &          \faCheckSquare                                                                                                                                                                                                                           & \hspace{-1em}\rdelim\}{4.5}{*}[\multirow{2}{*}{\begin{tabular}[c]{@{}l@{}}42 participants \\ (Random sampling: \\ 37 participants; purposive \\ sampling: 5 participants)\end{tabular}}] \\ \cmidrule(r){1-3} \cmidrule(lr){5-5}
DC3.2.    & \begin{tabular}[c]{@{}l@{}}Sending the survey link to our known \\ contacts in the software \\ development industry\end{tabular}                                                              &                       \faCheckSquare                                                                                                                    &                                                                                                    &                                                                                                                     \faCheckSquare                                                                                                                &                                                                                                                                                          \\ \cmidrule(r){1-3} \cmidrule(l){5-6} 
DC3.3.    & Using other recruitment techniques                                                                                                                                                                              & \begin{tabular}[c]{@{}l@{}}Agile Alliance posting the survey \\ link on their \\ L\textit{inkedIn, Twitter,} and \textit{Facebook} \\ channels\end{tabular} &                                                                                                    & \begin{tabular}[c]{@{}l@{}}Recruiting participants \\ through \textit{Amazon} \\ \textit{Mechanical Turk}\\ \\ Qualification criteria:\\ \tabitem Employment Industry - \\ Software and \\ IT Services\\ \tabitem Job Function - Information \\ Technology\end{tabular} & \begin{tabular}[c]{@{}l@{}}159 participants \\(Purposive sampling)   \end{tabular}                    \\  \midrule
 & \textbf{Collected data relavant to this paper} & \textbf{Quantitatve and Qualitative data} & & \textbf{Qualitative data} & 
\\ \bottomrule
\end{tabular}%
}
\end{table*}




\subsection{Data Analysis} \label{sec:data_analysis}
The data analysis steps we followed are given in Table \ref{tab:data_analysis}. Survey Alpha and Beta followed different approaches in data analysis.

\begin{table*}[]
\caption{Data Analysis}
\label{tab:data_analysis}
\resizebox{\textwidth}{!}{%
\begin{tabular}{@{}llll@{}}
\toprule
\textbf{} & \textbf{Step}                                         & \textbf{Survey Alpha}                                                                                                                                                                         & \textbf{Survey Beta}                                                                                                                                                                                                                            \\ \midrule
DA1.      & Quantitative analysis                                 & \begin{tabular}[c]{@{}l@{}}Descriptive statistical analysis: \\ Validation of pre-defined RC challenging factors\\ \\ \textbf{RQ answered: RQ1}\end{tabular}                                       & N/A to this paper                                                                                                                                                                                                                            \\ \midrule
DA2.      & Qualitative analysis                                  & \begin{tabular}[c]{@{}l@{}}Strauss--Corbinian GT data analysis techniques -- \\ open coding and constant comparison: Further RC \\ challenging factors, most common pre--development \\RC handling techniques\\ \\ \\ \\ \\ \textbf{RQ answered: RQ1, RQ4}\end{tabular} & \begin{tabular}[c]{@{}l@{}}DA2.1. JAWS: Emotion extraction from raw data\\ \\ DA2.2. STGT4DA -- open coding and constant \\ comparison: stakeholder factors of \\ emotional responses to RCs, emotional responses\\to RC challenging factors,\\most common pre--development \\RC handling technniques\\ \\ \textbf{RQ answered: RQ2, RQ3, RQ4}\end{tabular} \\ \midrule
DA3.      & \begin{tabular}[c]{@{}l@{}}Data synthesis of findings from \\survey Alpha and survey Beta\end{tabular} & \multicolumn{2}{l}{\begin{tabular}[c]{@{}l@{}}STGT4DA -- memoing through visual memos (diagrams): Conceptual model\\ \\ \textbf{The relationship between the findings}\end{tabular}}                                                                                                                                                                                                                                                                                   \\ \bottomrule
\end{tabular}%
}
\end{table*}

\textit{Quantitative Analysis (DA1):} Quantitative analysis where descriptive statistical analysis was done for quantitative data collected via both surveys. However, we only report the quantitative analysis of survey Alpha findings (using \textit{Microsoft Excel}) here, as survey Beta quantitative analysis findings (using \textit{Python}) are out of the scope of this paper,  and are presented in \cite{Madampe2022TheEngineering}. 

\textit{Qualitative Analysis (DA2):} Qualitative analysis approaches in the 2 surveys are different from each other -- survey Alpha focused only on RCs, and survey Beta focused on both emotions and RCs. 
Survey Alpha only followed the open coding and constant comparison techniques of Strauss--Corbinian Grounded Theory (GT) \cite{Strauss1998BasicsTechniques.} (using \textit{Microsoft Excel}). We used Strauss--Corbinian GT due to its structured approach of data analysis and our previous experience in using it. By the time of the first study, STGT4DA was not introduced yet. Therefore, STGT4DA was not used in survey Alpha.

In survey Beta, we followed a combined qualitative approach of JAWS analysis and open coding and constant comparison techniques as in STGT4DA \cite{Hoda2021Socio-TechnicalEngineeringb} (using MAXQDA\footnote{https://www.maxqda.com/} majorly and \textit{Microsoft Excel} as needed). We used STGT4DA due to its suitability to apply in socio--technical studies, its similarity to Strauss--Corbinian GT data analysis techniques where we had previous experience in,  and our interest in applying it.
Since the open coding and constant comparison techniques are the same in Strauss--Corbinian GT and STGT4DA, here we only explain the combined qualitative data analysis approach we took in Survey Beta.  Below, we further explain the analysis through examples.


\textit{DA2.1. JAWS analysis:} As the open--ended questions were developed to allow participants to report their experiences in feeling the specific emotions as given in the question (emotion sub--scale of JAWS), the participants used the exact terms of emotions as in JAWS. In cases where the participants did not mention the emotion, we considered that they felt the emotions in that particular sub--scale in general. For example, we extracted the emotion ``anxious'' from raw data ````\textit{Anxious when I feel the new technology is difficult to learn}''.

\textit{DA2.2. STGT analysis:} STGT served our analysis in two key ways. (1) It helped us in identifying the evidence in survey Beta data for emotions for RC handling challenging factors found in survey Alpha. As given in Table \ref{tab:challenges_emotions}, we found evidence for each RC challenging factor along their dimensions using STGT4DA. For example, when participant $\beta$P94 answered the question on when he felt high\textsuperscript{2} emotions as ``\textit{when the implementation turns out to be less complex than initially specified}'', we were able to know that when complexity (challenging factor) is low (dimension), $\beta$P94 felt high\textsuperscript{2} emotions in general. However, we found these organically, i.e., we did not force the findings to emerge but later on aided in identifying evidence for RC challenging factors in survey Beta. (2) emergence of pre--development techniques and stakeholder factors in survey Beta. 

We first open coded the qualitative data (interpreted the data in small, meaningful chunks of words). We then compared these codes using constant comparison and produced concepts where similar codes were grouped together. Constant comparison was applied again on concepts to produce sub--categories. After sub--category creation, we repeated the same to produce categories. For example, raw data ``\textit{Anxious when I feel the new technology is difficult to learn}'' yielded in the code ``difficulty in learning [new technology]'' which we compared with similar codes to produce the concept ``difficulty in learning/ acquiring new knowledge’, followed by the sub-category ``practitioner--related factors'' and the category ``stakeholder factors''.

\textit{DA3. Data synthesis:} Memoing can take different forms (verbal, textual, visual). Here we used visual memos as it was easy to uncover the connections between the findings of the two studies. We observed the findings from a socio--technical perspective, and that allowed us to identify the connection between the findings. We found the challenging factors of effort, cross-functionality within the team, and access to the customer related to the stakeholders. The conceptual model representing the relationships between the findings from survey Alpha and survey Beta is given in Fig. \ref{fig:relationship}.

\section{Participant and Project Context}

\subsection{Demographic Data of Participants}
A summary of demographic data of participants for both surveys is given in Table \ref{tab:demo}. The majority of survey Alpha participants represented Asia (N=26; 65\%) whereas the majority of survey Beta participants represented North America (N=96; 47.78\%). The most commonly played role of both surveys' participants was developer (Alpha (N=18; 45\%); Beta (N=75; 37.31\%)). Survey Alpha participants had a mean total experience of 8.74 years (min(total experience)=1 year; max(total experience)=30 years), and a mean total agile experience of 4.4 years (min(total agile experience)=1 year; max(total experience)=20 years). Survey Beta participants had a mean total experience of 7.8 years (min(total experience)=1 year; max(total experience)=35 years), and a mean total agile experience of 5.12 years (min(total agile experience)=0 years; max(total experience)=20 years). 

\begin{table}[]
\caption{Demographic Information of Survey Participants (Dev: Developer; AC/SM: Agile Coach/Scrum Master; BA: Business Analyst; PO: Product Owner; XT: Total Software Development Experience; XTA: Total Agile Experience)}
\label{tab:demo}
\resizebox{\columnwidth}{!}{%
\begin{tabular}{@{}llll@{}}
\toprule

\multicolumn{4}{l}{\cellcolor[HTML]{EFEFEF}\textbf{Survey Alpha} (Other: $\leq$ 2 participants)}                                                              \\ \midrule
\textbf{Location} & \textbf{\# of Participants} & \textbf{Role} & \textbf{\# of Participants} \\ \midrule
Asia              & \mybarhhigh{2.6}{26}                          & Dev           & \mybarhhigh{1.5}{15}                          \\
Australasia       & \mybarhhigh{0.9}{9}                           & Tester        & \mybarhhigh{0.8}{8}                           \\
North America     & \mybarhhigh{0.3}{3}                           & AC/SM         & \mybarhhigh{0.5}{5}                           \\
Europe            & \mybarhhigh{0.2}{2}                           & BA            & \mybarhhigh{0.3}{3}                           \\ \cmidrule(r){1-2}
\textbf{Gender}            & \textbf{\# of Participants}          & PO            & \mybarhhigh{0.3}{3}                           \\ \cmidrule(r){1-2}
Male              & \mybarhhigh{2.3}{23}                          & Other         & \mybarhhigh{0.6}{6}                           \\
Female            & \mybarhhigh{1.7}{17}                          &               &                             \\ \midrule
\textbf{XT}                & \textbf{\# of Years}                 & \textbf{XTA}           & \textbf{\# of Years}                 \\ \midrule
Minimum           & 1                           & Minimum       & 1                           \\
Maximum           & 30                          & Maximum       & 20                          \\
Mean              & 8.74                        & Mean          & 4.4                         \\  \midrule

\multicolumn{4}{l}{\cellcolor[HTML]{EFEFEF}\textbf{Survey Beta} (Other: $\leq$ 5 participants)}                                                              \\ \midrule

\textbf{Location} & \textbf{\# of Participants} & \textbf{Role} & \textbf{\# of Participants} \\ \midrule
North America     & \mybarhhighY{19.2}{96}                          & Dev           & \mybarhhighY{15}{75}                          \\
Asia              & \mybarhhighY{8}{40}                          & Manager       & \mybarhhighY{4.2}{21}                          \\
Europe            & \mybarhhighY{4.8}{24}                          & BA            & \mybarhhighY{3.8}{19}                          \\
Australasia       & \mybarhhighY{4.4}{22}                          & Dev, Tester   & \mybarhhighY{2.8}{14}                          \\
South America     & \mybarhhighY{3.4}{17}                          & Tester        & \mybarhhighY{2}{10}                          \\
Africa            & \mybarhhighY{0.4}{2}                           & Dev, Manager  & \mybarhhighY{1.8}{9}                           \\ \cmidrule(r){1-2}
\textbf{Gender}   & \textbf{\# of Participants} & AC/SM         & \mybarhhighY{1.6}{8}                           \\ \cmidrule(r){1-2}
Male              & \mybarhhighY{23}{115}                         & AC/SM, Dev    & \mybarhhighY{1.4}{7}                           \\
Female            & \mybarhhighY{17}{85}                        & Other         & \mybarhhighY{7.6}{38}                          \\
Gender diverse    & \mybarhhigh{0.2}{1}                           &               &                             \\ \midrule
\textbf{XT}       & \textbf{\# of Years}        & \textbf{XTA}  & \textbf{\# of Years}        \\ \midrule
Minimum           & 1                           & Minimum       & 0                           \\
Maximum           & 35                          & Maximum       & 20                          \\
Mean              & 7.84                        & Mean          & 5.12                        \\

\bottomrule
\end{tabular}%
}
\end{table}

\subsection{Project and Team Information of Participants}


A summary of project and team information for both surveys is given in Appendix A.
The projects that both survey participants chose to answer the questionnaires were new developments (Alpha (N=26; 65\%), Beta (N=115; 57.21\%)). All participants of survey Alpha used agile methods in their projects (N=40; 100\%) as we only targeted agile practitioners in survey Alpha, and the majority of survey Beta participants used agile methods in their projects as well (N=176; 87.56\%). Therefore, overall, 89.32\% (N=216) participants from both the surveys used agile in their projects which is in line with reported agile use in the industry \cite{202014thAgile}, where RCs are common.

\section{Findings} \label{sec:findings}
\subsection{Factors that make RC handling a challenge (Answer to RQ1)}
\label{sec:challenge}

\begin{figure}
    \centering
    \includegraphics[width=\columnwidth]{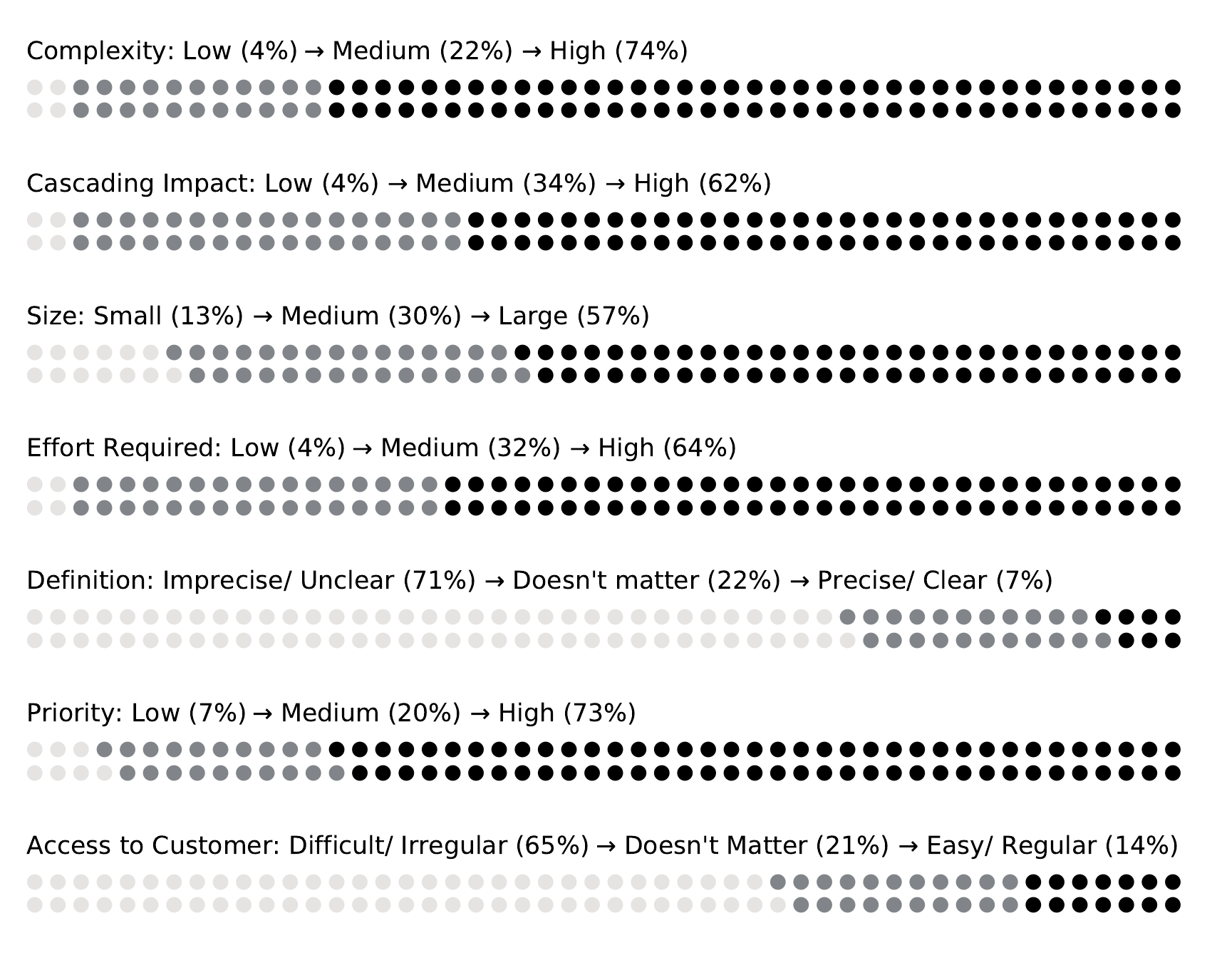}
    \caption{Factors Making Requirements Changes Challenging to Handle: Results from Survey Alpha Quantitative Data}
    \label{fig:quan}
\end{figure}




Quantitative analysis of survey responses are shown in Fig. \ref{fig:quan}. Taking the top most responded results by the participants into consideration, an RC is seen as challenging when its
\begin{itemize}
    \item \textbf{complexity} is \textbf{high} and/or;
    \item \textbf{cascading impact} is \textbf{high} and/or;
    \item \textbf{size} is \textbf{large} and/or;
    \item \textbf{effort required} is \textbf{high} and/or;
    \item \textbf{definition} is \textbf{imprecise or unclear} and/or;
    \item \textbf{priority} is \textbf{high} and/or;
    \item \textbf{access to customer} is \textbf{difficult or irregular}.
\end{itemize}

Apart from complexity, cascading impact, size, effort required, definition, priority, and access to customer, our  analysis of open--ended responses resulted in identifying that \textit{forced cross-functionality} make the RC handling challenging. For example, when business analysts try to force the completion of development; there is insufficient impact analysis of the RC by developers; and developers are disengaged from thinking deeply about the RC.
     
    \begin{quotation}
    \textit{``..disengaged from thinking \& expecting others to do the thinking \& exploring of expected business value (“I just want to write code, man!”)'' -- $\alpha$P25}
    \end{quotation}
    
Based on the above quantitative and qualitative data analysis, we define a challenging RC as per below:

\begin{tcolorbox}[colback=gray!10, boxrule=0pt,frame hidden]
\textbf{The definition of a challenging RC.} An RC whose complexity is high, cascading impact is high, size is large, effort required to action is high, definition is imprecise or unclear, priority is high, access to customers is difficult or irregular, and cross--functionality is forced is called a \textit{\textbf{challenging RC}}.
\end{tcolorbox}

\subsection{How do Practitioners Feel when it is Challenging to Handle RCs (Answer to RQ2)}
The key challenge factors, the key emotions felt, some representative examples, and roles of the participants who reported the challenges and emotions are given in Appendix B and described below.

\textbf{Complexity.}
As stated by several participants, when the RC is not complicated to work on, they tend to feel both high\textsuperscript{2} (e.g.: \textit{excited, energetic}) and high\textsuperscript{1} (e.g.: \textit{content, calm, relaxed}) emotions. Even when the RC complexity that was originally specified changes to a less complicated one, they tend to feel high pleasurable emotions. For example, as $\beta$P194 mentioned, they felt high\textsuperscript{2} emotions when the implementation turned out to be less complex than it was initially specified.
In contrast, when an RC is complicated to work on, they tend to feel more low\textsuperscript{1} emotions (e.g.: \textit{anxiety}), as mentioned by a few participants. For example, $\beta$82 mentioned, additions and/or modifications that were highly complex made them feel low\textsuperscript{1} emotions.

\textbf{Cascading Impact.}
When the impact of the RC on the other requirements, including the ones that have already been developed, is low, the practitioners tend to feel high\textsuperscript{1} emotions (e.g.: \textit{content}). However, when the cascading impact is high -- such as when several changes have to be done together to realise the RC, such as design-level changes in the database, application, and program structure -- practitioners tend to feel both low\textsuperscript{1} (e.g.: \textit{angry}) and low\textsuperscript{2} (e.g.: \textit{discouraged, fatigued}) emotions.

\textbf{Size of the RC.}
When the size of the RC is large, the practitioners tend to feel low\textsuperscript{2} emotions in general. For example, as $\beta$P10 mentioned, when RC additions such as new features are large, it makes them rewrite on a large scale, and also goes beyond the initial scope. This indicates that the size of the RC, impacts effort (rewriting incurs effort) and increases the scope -- i.e., scope creep. Not only the project scope, but also the scope of work of the individual is impacted. For example, when the RCs are within scope, practitioners tend to feel high\textsuperscript{2} emotions (e.g.: \textit{enthusiastic}).
Even though we found evidence of emotional responses felt towards large-sized RCs, we could not find how practitioners feel when their RCs are small.

\textbf{Definition.}
Precise and clear definition of an RC matters. When the RC is well-defined, practitioners tend to feel high\textsuperscript{1} emotions in general. For example, $\beta$P175 mentioned that, all required details for working on the RC were well specified and that made them know what exactly they were supposed to do and that did not require them to do a second pass on the code for checking the consistency with the rest of the UI work.
On the other hand, when RCs are not well-defined, practitioners tend to feel low\textsuperscript{2} emotions. Practitioners also struggle to look for information needed to work on the RC, which is connected to the challenge ``access to customer''. 

\textbf{Access to Customer.}
When practitioners have regular conversations about an RC and engage with their team throughout the process, they tend to   feel high\textsuperscript{1} emotions (e.g.: \textit{calm}).
However, some practitioners stated that when the customer is busy and there are delays in replying about the RC, does not read the emails fully, or even not taking the meetings seriously, they feel low\textsuperscript{2} emotions (e.g.: \textit{fatigued, gloomy}).
This is related to the relationship management of the customer that we explain in the next section.

We did not find specific reported emotional responses for RC priority and cross--functionality aspects. 

From the above findings, the variations of emotions found across dimensions of RC challenging factors are summarised below.
\begin{tcolorbox}[colback=gray!10, boxrule=0pt,frame hidden]

\textbf{Variations of emotions across dimensions of RC challenging factors.} 
When an RC is complicated, practitioners feel low\textsuperscript{1} emotions. When an RC is less complicated, practitioners feel both high\textsuperscript{2} and high\textsuperscript{1} emotions. When the cascading impact of the RC is high, practitioners tend to feel both low\textsubscript{2} and low\textsuperscript{1} emotions. When the cascading impact is low, the emotions felt by the practitioners tend to be high\textsuperscript{1}. Low\textsuperscript{2} emotions tend to be felt when an RC is large. When the definition of the RC is precise/ clear, high\textsuperscript{1} emotions are felt. However, low\textsuperscript{2} emotions are felt when the RC is defined imprecisely or unclearly. When the access to the customer is easy/ regular, calmness is often felt by the practitioners. When it is difficult/ irregular to access the customer, emotions such as fatigue and gloom are felt. Further, when the effort required to work on the RC is high, low\textsuperscript{2} emotions tend to be felt by practitioners.
\end{tcolorbox}


\subsection{How Stakeholders Influence the Overall Handling of the RCs (Answer to RQ3)}
\label{sec:stakeholder}


We found several stakeholder factors that are perceived by the practitioners as factors that influence the overall RC handling and resulting in their emotional responses. The key stakeholders found are the ``practitioner'', ``team'', ``manager'', and ``customer''. The influence of each of these stakeholders is described in each of the sub--sections below. We describe the key factors, the felt emotions, some representative examples, and the roles of the participants who reported them. We summarise the factors that lead to high and low pleasurable emotions of the practitioners, that we term ``high pleasurable and low pleasurable factors". Some factors take both high and low pleasurable dimensions -- for example, learning/ knowledge acquisition arouses high pleasurable emotions when that is preferred by the practitioners. However, when it is difficult to learn/ acquire the knowledge, low pleasurable emotions may be aroused. 
For such cases, we present the factors as single description points.






The factors not only result in arousing the emotions of the practitioners but may also have work--related consequences. We have indicated these work--related consequences by \textit{PC\textless ID\textgreater} for the practitioner, \textit{TC\textless ID\textgreater} for the team, \textit{MC\textless ID\textgreater} for the manager, and \textit{CC\textless ID\textgreater} for the customer.

\subsubsection{How Practitioners Themselves Influence the Handling of RCs (Answer to RQ3.1.)}
Below, we describe the practitioner--related factors. Representative examples for these factors are given in Appendix C.

\begin{tcolorbox}[colback=gray!10, boxrule=0pt,frame hidden]
\textbf{High pleasurable practitioner--related factors:}\\
Preference in learning/ knowledge acquisition {\footnotesize (\faBrain A)}\\
High empowerment/autonomy/ responsibility {\footnotesize (\faBrain B)}\\
Ideating {\footnotesize (\faBrain C)}\\

\textbf{Low pleasurable practitioner--related factors:}\\
Difficulty in learning/ knowledge acquisition {\footnotesize (\faBrain A)}\\
Less autonomy {\footnotesize (\faBrain B)}\\
Making mistakes {\footnotesize (\faBrain D)} 
\end{tcolorbox}

\textbf{\faBrain A. Learning/ knowledge acquisition of the practitioner.}
Some practitioners prefer acquiring new knowledge and gaining a learning experience through handling an RC. They see knowledge acquisition and learning as an antecedent that makes them feel high\textsuperscript{2} emotions (e.g.: \textit{excited, inspired}), which also improves their skills (PC1), allows them to feel ``smarter'' (PC2), and let the practitioners meet peers’ requests if they ask for any (PC3).
However, learning and knowledge acquisition is not easy for everyone. There are cases where practitioners find it difficult to learn-- for example, learning a new technology, a tool or a new methodology. In such cases, the practitioners may experience low\textsuperscript{1} (e.g.: \textit{anxious}) and low\textsuperscript{2} (e.g.: \textit{gloomy, fatigued, discouraged, over depressed}) emotions.

\textbf{\faBrain B. Empowerment/ Autonomy/ Responsibility of the practitioner.}
The practitioners feeling empowered, having autonomy, and feeling responsible while working on RCs result in high\textsuperscript{2} (e.g.: \textit{energetic, inspired}) and high\textsuperscript{1} (e.g.: \textit{satisfied, relaxed}) emotions within them as they reported. 
According to self--determination theory, autonomy is one of the psychological needs of human beings. When they are being controlled, their autonomy is restricted, hence the psychological need is violated. Several participants reported that they are required to work overtime because of the RCs. This forced effort often leads to low\textsuperscript{1} emtions and low\textsuperscript{2} (e.g.: \textit{bored, depressed}) emotions, especially in \textit{depression}, as perceived by the practitioners. Not only low pleasurable emotions, but also working overtime leads to losing sleep (PC4) and time for relaxation (PC5), which are basic physiological needs as in Maslow’s hierarchy of human needs.

\textbf{\faBrain C. Practitioner ideating.}
Practitioners coming up with ideas to work on the RC, often generate high\textsuperscript{2} emotions in general in them as they reported. Even the ideas that practitioners get at times when they use ideating as an exercise for their brains, become useful when they face a similar occasion in real, in this case a similar RC.

\textbf{\faBrain D. Practitioner making mistakes.}
Practitioners themselves making minor mistakes and finding that their mistakes could have been avoided through easy fixes, make them feel both low\textsuperscript{1} in general and low\textsuperscript{2} (e.g.: \textit{over depressed}) emotions.

\subsubsection{How Peers in the Team Influence the Handling of RCs (Answer to RQ3.2)}
Below, we describe the peer--related factors. Representative examples for these factors are given in Appendix D.

\begin{tcolorbox}[colback=gray!10, boxrule=0pt,frame hidden]
\textbf{High pleasurable team--related factors:}\\
Having the required expertise/ competency/ relevant experience within the team {\footnotesize(\faUsers A)}\\
High efficiency/ productivity of the team {\footnotesize(\faUsers B)}\\
Better relationship management within the team through collaboration and engagement {\footnotesize(\faUsers C)}\\

\textbf{Low pleasurable team--related factors:}\\
Low efficiency/ low productivity of the team {\footnotesize(\faUsers B)}\\
Team members making mistakes {\footnotesize(\faUsers D)}
\end{tcolorbox}

\textbf{\faUsers A. Having the required expertise/ competency/ relevant experience within the team.}
Having team members who are experts, competent, and having relevant experience in the team tend to arise high\textsuperscript{2} (e.g.: \textit{inspired}) and high\textsuperscript{1} (e.g.: \textit{relaxed, satisfied, calm, at ease}) emotions in practitioners when they are handling RCs. For example, our participants mentioned that peers with the necessary knowledge, make them inspired and peers having experience working with similar RCs allow the practitioners to stay \textit{calm, relaxed, satisfied, at ease} when working with RCs.

\textbf{\faUsers B. Efficiency/ Productivity of the team.}
The effect of efficiency and productivity of peers is two--fold. Being efficient and productive tends to lead to high pleasurable emotions, and being inefficient and unproductive tends to lead to low pleasurable emotions of practitioners when they are working on RCs.
When the team has members who are efficient and productive, this results in both high\textsuperscript{2} (e.g.: \textit{inspired, excited, energetic}) and high\textsuperscript{1} (e.g.: \textit{relaxed, satisfied, calm, at ease}) emotions of  practitioners, as a better outcome is anticipated. At the same time, when the team is efficient and productive, the absence of a manager does not make much difference, as mentioned by $\beta$P193 who is both a developer and a manager.

On the contrary, team members whose rate of working is slow, and who believe in traditional ways of executing tasks, tend to result in low efficiency and low productivity, and low\textsuperscript{1} (e.g.: \textit{angry, anxious}) emotions in practitioners, as perceived by them. Also, when team members do not work as expected by practitioners, that also may lead to low\textsuperscript{1} emotions. 
$\beta$P156 claimed that they noticed that slow working rates are sometimes apparent in practitioners who are older in age, and also such practitioners may show a resistance to adapt to current practices at work. In relation to this, some participants stated that they feel high pleasurable emotions when working with young peers. These potentially controversial statements related to team member age differences requires further research in the future,  and we elaborate this in Section \ref{imp:researchers}.



\begin{quotation}
\textit{``I feel angry, anxious anytime I work with old folks who always have a slow work rate and always believes in their traditional way of performing duties I am always furious and always hope I get younger teammates'' -- $\beta$P156} 

\textit{``Also because of this two young individual added to our team, I feel calm, mainly because I know all work will be done perfectly'' -- $\beta$P156}

\textit{``I have young and vibrant team members who always find new ways to sort things out whenever we deal with software and I am always energetic and excited to work with this particular set of people'' -- $\beta$P60} 
\end{quotation}

\textbf{\faUsers C. Relationship management within the team.}
Good working relationships, where collaborative working and team engagement play major roles, are essential to effectively handling RCs. When team members work effectively, being accountable for what they do, and ask for and receive help from others in the team, this tends to make individual practitioners feel both high\textsuperscript{2} (e.g.: \textit{inspired, energetic}) and high\textsuperscript{1} (e.g.: \textit{relaxed, satisfied, calm}) emotions. Having supportive peers also result in the efficiency and productivity of the team (TC1).
Better collaboration and team engagement also enable finding solutions easily (TC2), can set a practitioners' mood to a ``jolly'' level (TC3), lessens the probability of feeling frustrated and pressurised (TC4), achieve the goal/ objective on time (TC5), and stay within the budget (TC6).

When a team works well, and management communicates well with the team, and most importantly when the team works collaboratively and engages well by conducting meetings to work on the RCs, the high\textsuperscript{2} (e.g.: \textit{inspired, energetic}) and high\textsuperscript{1} (e.g.: \textit{relaxed, satisfied, calm}) emotions of practitioners arise as reported by our participants. For example, $\beta$P3 stated that them having a meeting to discuss the RC properly, and to break down the RC into tasks, prioritise them, estimate effort, and allocate the tasks within their working hours led to high pleasurable emotions in him. 

\textbf{\faUsers D. Team members making mistakes.}
When peers make mistakes, but not the practitioners themselves, they may feel low\textsuperscript{1} (e.g.: \textit{angry, disgusted}) emotions in general, and specifically anger and even disgust. They also believe that not only themselves, but also others in the team feel the same ways when their peers make mistakes.

\subsubsection{How Managers Influence the Handling of RCs (Answer to RQ3.3.)}
Below, we describe the manager--related factors. Representative examples for these factors are given in Appendix E.

\begin{tcolorbox}[colback=gray!10, boxrule=0pt,frame hidden]
\textbf{High pleasurable manager--related factors:}\\
A manager having empathy (social awareness) {\footnotesize (\faUserSecret A)}\\
A manager building better relationships through communication and coordination (relationship management) {\footnotesize (\faUserSecret B)}\\

\textbf{Low pleasurable manager--related factor:}\\
A manager lacking empathy (social awareness) {\footnotesize (\faUserSecret A)}
\end{tcolorbox}

\textbf{\faUserSecret A. Social awareness of the manager.}
Both positive and negative aspects of social awareness of the manager --  having empathy or lacking it -- are perceived as factors that lead to the high and low pleasurable emotions of the practitioners. 
For example, when a manager does not pressurise their team, but motivates the team (MC1) by being empathetic and allowing the team to handle their RCs as best suits them, practitioners tend to experience an arising of high\textsuperscript{2} emotions (e.g.: \textit{energetic}), and high\textsuperscript{1} (e.g.: \textit{calm}) emotions of the practitioners.
On the other hand, when a manager lacks empathy e.g. when they are unable to cope with human errors of the team -- even minor mistakes -- and when the manager does not seem to feel how the team feels or understand the team’s emotional investments in their RCs, practitioners perceive that it leads to having both low\textsuperscript{2} in general and low\textsuperscript{1} (e.g.: \textit{angry}) emotions in them.

\textbf{\faUserSecret B. Relationship management of the manager.}
Manager’s prompt and honest communication, better coordination, and having conversations with the team tend to make the practitioners feel both high\textsuperscript{2} (e.g.: \textit{inspired}) and high\textsuperscript{1} (e.g.: satisfied, relaxed) emotions that allows them to better handle the RCs.

\subsubsection{How Customers Influence the Handling of RCs (Answer to RQ3.4.)}
Below, we describe the customer--related factors. Representative examples for these factors are given in Appendix F.

\begin{tcolorbox}[colback=gray!10, boxrule=0pt,frame hidden]
\textbf{High pleasurable customer--related factors:}\\
Customer’s manifested high pleasurable emotions (self--regulation) {\footnotesize (\faPeopleArrows A)}\\
Customer’s positive engagement with the team (relationship management) {\footnotesize (\faPeopleArrows B)}\\

\textbf{Low pleasurable customer--related factors:}\\
Customer’s manifested low pleasurable emotions (self--regulation) {\footnotesize (\faPeopleArrows A)}\\
Customer’s scepticism of what he/she/they need {\footnotesize (\faPeopleArrows C)}
\end{tcolorbox}

\textbf{\faPeopleArrows A. Self--regulation of the customer.}
A customer manifesting their emotions has an impact on the emotions of the practitioners. For example, when the customer is satisfied or excited with practitioners’ work -- that is when the customer manifests both high\textsuperscript{2} (e.g.: \textit{excited}) and high\textsuperscript{1} (e.g.: \textit{satisfied}) emotions -- the practitioners tend also to feel both high\textsuperscript{2} (e.g.: \textit{energetic, inspired}) and high\textsuperscript{1} (e.g.: \textit{calm, satisfied}) emotions as well. 
Likewise, if the customer appears to be unsatisfied or unimpressed with the delivered work -- that is, when the manifested emotions of the customer are low pleasurable emotions (e.g.: \textit{unsatisfied, unimpressed}) -- then the practitioners tend to feel angry and low\textsuperscript{2} (e.g.: \textit{depressed}) emotions. As a further consequence of this, the team may have to redo some RC work (CC2) as well.

\textbf{\faPeopleArrows B. Relationship management of the customer.}
Customer’s positive engagement with the project and the team when they are handling RCs leads to high\textsuperscript{2} and high\textsuperscript{1} (e.g.: \textit{calm, satisfied}) emotions of the practitioners. For instance, having a better understanding of the RC and customer working together with the team, and giving input to the team leads to high pleasurable emotions of the practitioners. Customer’s positive engagement not only leads to high pleasure emotions of the team but also allows defining the RC, thereby resulting in a better outcome.

\textbf{\faPeopleArrows C. Customer's scepticism of what they need.}
If practitioners perceive that their customer is not clear about what they need and request RCs repeatedly, then practitioners perceive that as an antecedent that makes them feel both low\textsuperscript{1} (e.g.: \textit{anxious}) and low\textsuperscript{2} (e.g.: \textit{bored, depressed, discouraged}) emotions. This also possibly leads to the team redoing their RC work (CC2) and the team not fully addressing the RC, expecting that it will change again (CC3). Furthermore, Hoda et al. \cite{Hoda2011TheTeams} mentions that customer's scepticism leads to inadequate collaboration during software development.




\section{How do practitioners approach handling RCs while managing emotions? (Answer to RQ4)}
Some common RC handling techniques were reported by practitioners -- ranging from processes including high-level practices to low-level coding techniques -- that they use and that make their RC handling more emotionally easier. We list them below for the benefit of other practitioners. 













\textit{Re-estimate and change the sprint plan (from Survey Alpha).} Some practitioners reported the need to periodically revisit RC change handling priorities and ordering, to help them manage negative reactions. This includes analysing the impact of the RC $\rightarrow$ estimating the effort of the RC $\rightarrow$ adding RC to the iteration backlog $\rightarrow$ removing existing user stories according to the priority and size from the iteration backlog.

\textbf{RC Prioritisation (from Survey Alpha).} As mentioned by several participants, prioritisation can be done based on the changes in the market, based on team leaning, and the business need of the customer, or combinations. 
However, if the customer is unclear or sceptical about what they need, then it is the responsibility of the manager and the team to move forward with a discussion which resolves different RC priorities.

\textbf{Discuss with the project manager whether the RC is reasonable to implement (from Survey Beta).} Practitioners suggest $\rightarrow$ identify the scope affected $\rightarrow$ discuss with the relevant product owner $\rightarrow$ quick POC implementation $\rightarrow$ plan $\rightarrow$ design $\rightarrow$ assign work. This indicates the utilisation of continuous relationship management before making any decision on the implementation/ acceptance of the RC to help manage practitioner emotions.

\textbf{First come, first served (from Survey Alpha and Beta).} In some cases, participants implement RCs as they receive them. We identified this phenomenon in one of our previous studies as well \cite{Madampe2020TowardsTeamsb}. This technique does not follow any prioritisation or any other best practices, but simply implementing them as received. Sometimes this strategy works well for managing emotional impact of RCs e.g.

\begin{quotation}
\textit{``This is my regular demeanor while working.  I try to deal with things as they come'' -- $\beta$P52}
\end{quotation}


\textbf{Find your own ways to handle the RC and use them in similar situations (from Survey Beta).} Sometimes the so--called standard ways of handling RCs may not be applicable to every situation. In such instances, practitioners and their teams need to find ways that suit their own situation, team members and customers and apply them. For example, $\beta$P56 mentioned that having these in hand could be useful in applying at similar situations:

\begin{quotation}
\textit{``I feel and have felt this way whenever we found a better way of doing things.  One method might work or it might not but then suddenly, we find a way to push requests faster, anything like that and it gets me really excited and hopeful for the project.  I feel like there's other places I could apply these things to'' -- $\beta$P56}
\end{quotation}

\textbf{Write pseudocode to generate solutions (from Survey Beta).} Before implementing an RC, working through design changes and pseudocode was suggested as an effective technique by some participants, allowing them to gauge impact on emotions up front before fully implementing an RC.

\textbf{Early implementation of a flexible structure to manage the logic flow (from Survey Beta).} This technique is useful for handling cascading impacts of an RC. Participant $\beta$P122 mentioned that the flexible structure they had where they used dependency injection help them to more easily implement many RCs:

\begin{quotation}
\textit{``In one of my previous projects, my leader and project manager wanted to adjust a piece of the predefined logic flow, because I had already implemented a flexible structure to manage that logic flow, I just need to change the order of parameters and then everything work to fit the new requirement.
Emotion: at ease.
Process: Scrum.
Techniques used to handle changes: program to interfaces, dependency injection'' -- $\beta$P122}
\end{quotation}




The above results demonstrate that some RC handling ways are highly organised where a step-by-step process is followed, and some are not. However, common techniques of RC handling include, but are not limited to, impact analysis, prioritisation, effort estimation, and POC implementation or writing pseudocode before implementation. 

\section{Discussion}

\subsection{Relationship between Challenges, Stakeholders, and Emotions}

\begin{figure*}[b]
    \centering
    \includegraphics[width=\textwidth]{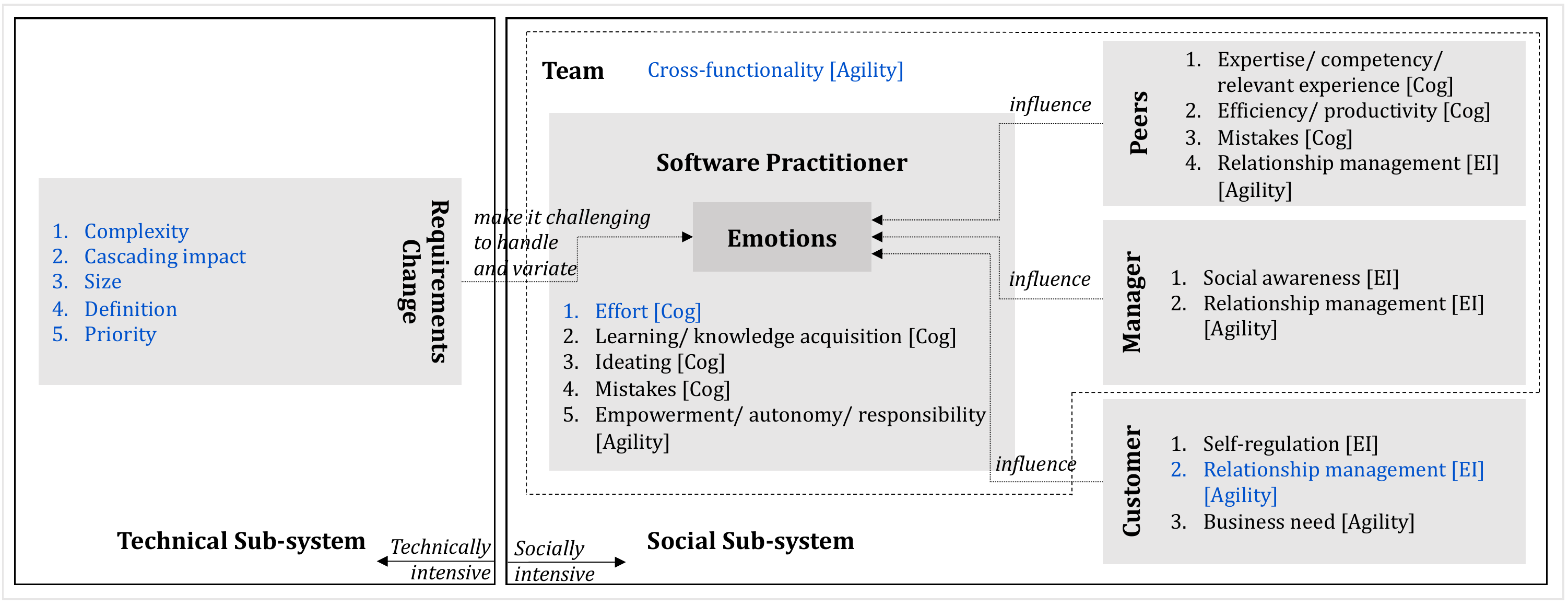}
    \caption{Relationship between the Challenging Factors, Stakeholders, and Emotions (Cog: Cognitive Intelligence; EI: Emotional Intelligence $\vert$ Blue Text: Challenging Factors; Black Text: Stakeholder Factors)}
    \label{fig:relationship}
\end{figure*}




\textbf{Challenge factors and stakeholder factors.} Some of the challenging factors of RC handling reported and their impact on practitioner emotions -- complexity, cascading impact, size, definition, and priority -- are more technical aspects of handling RCs.Effort, cross--functionality, access to the customer, manager impact, and relationship management of the customer are socio-technical aspects of handling RCs. Fig. \ref{fig:relationship} summarises the key relationships between these factors that we have identified. The challenging factors are given in blue text and stakeholder factors are given in black text. 

\textbf{Challenging factors, stakeholders, and emotions.} From our findings, it is evident that RC handling challenging factors and stakeholders factors both contribute to the triggering of emotions in practitioners. Previously \cite{Madampe2022TheEngineering} we found that RCs and the stakeholders act as the stimuli of triggering emotions particularly at distinct events in the project and RC handling life cycles.

\textbf{Agility, emotional intelligence, and cognitive intelligence.} Further analysis of our findings resulted in categorising socio-technical factors into agility, emotional intelligence, and cognitive intelligence of practitioners and stakeholders. The majority of the factors under practitioner and peers belong to cognitive intelligence areas, whereas the majority of factors under customer and manager belong to relationship management which comes under emotional intelligence. In addition, cross--functionality, which is often a characteristic of agile teams, applies to practitioners, their peers, and manager. These categorisations are indicated next to the factors in Fig. \ref{fig:relationship}.

\subsection{Implications for Practitioners}






\label{imp:prac}

\textbf{RC challenge assessment as a pre-development RC handling technique. } Practitioners told us that there is a need to assess the challenging nature of RCs earlier -- i.e., at the receiving stage. They said that this would help practitioners to better understand how challenging the RC is to handle and ease the emotional impact during the RC handling process. The key challenging factors of RCs summarised in Section \ref{sec:challenge} and their respective dimensions can be used to assess how challenging RCs are (as challenge assessment metrics) before developing the RC.




\textbf{The interplay between agility, emotional intelligence, and cognitive intelligence.} From the relationship we identified between the challenging factors, stakeholders, and emotions (Fig. \ref{fig:relationship}), it is evident that for practitioners to better handle RCs while maintaining high pleasurable emotions in them (regulating low pleasurable emotions by navigating low pleasurable emotions towards high pleasurable emotions), the three aspects -- agility, 
emotional intelligence, 
and cognitive intelligence 
should be present at a sufficient level throughout the RC handling life cycle. 



\textbf{Emotional Awareness throughout RC Handling.}
Our findings indicate that practitioners feel various emotions while handling RCs, and as a final note we would like to emphasise the importance of emotions at work. Therefore, we request organisations, managers, and practitioners to, \textbf{make being aware of the emotional well--being a consistent and continuous practice.} Do not wait till something happens. 
Tools such as \textit{Emotimonitor} \cite{El-Migid2022Emotimonitor:Teams} can be used to monitor emotions at task level.

\begin{quote}
\textit{``Sometimes I feel depressed in bad working members. Many employers take an ad hoc approach to handling depression among employees. Many managers become aware of mental health issues only when they investigate why a team member is performing poorly.'' -- $\beta$P107
}
\end{quote}

\textbf{Recommendations.} Based on the above implications, in Table \ref{tab:recommendations} we provide best practices for practitioners, managers, and customers to consider following when handling RCs.

\begin{table*}[]
\caption{Recommendations}
\label{tab:recommendations}
\resizebox{\textwidth}{!}{%
\begin{tabular}{@{}lll@{}}
\toprule
\textbf{For}                                                                                          & \textbf{Recommendation}                                                                                                     & \textbf{How}                                                                                                                                                                                                                                                                                \\ \midrule
\multirow{7}{*}{\begin{tabular}[c]{@{}l@{}}Team including managers and \\ practitioners\end{tabular}} & \multirow{3}{*}{Improve agility}                                                                                            & Maintain a natural cross-functionality -- do not over do it by forcing                                                                                                                                                                                                                      \\ \\
                                                                                                      &                                                                                                                             & \begin{tabular}[c]{@{}l@{}}Have proper acceptance criteria for RC acceptance that suits you, and then follow them \\ to decide the acceptance of the RC. This may include questioning the rationality of the \\ RC, conducting impact analysis, estimating the effort required\end{tabular} \\ \\
                                                                                                      &                                                                                                                             & Prioritise the RC by discussing with the customer, manager, and the team                                                                                          \\ \cmidrule(l){2-3}                                                                            &                                        \multirow{4}{*}{\begin{tabular}[c]{@{}l@{}}Improve emotional intelligence \\ (self awareness)\end{tabular}}  \\ \\   &      
                                                                                                    & 
                                                                                                    \begin{tabular}[c]{@{}l@{}} Use emotion monitoring and tracking tools through out RC handling\end{tabular}

                                                                                                      \\ \midrule
\multirow{10}{*}{Practitioner}                                                                        & \multirow{4}{*}{\begin{tabular}[c]{@{}l@{}}Pre-development \\ RC handling techniques\end{tabular}}                          & \begin{tabular}[c]{@{}l@{}}Assess how challenging the RC is
\end{tabular}                                                                                                                         \\ \\
                                                                                                      &                                                                                                                             & \begin{tabular}[c]{@{}l@{}}Always discuss with the customer, manager, and the team early to clarify the necessary \\ information, better define the RC before implementing\end{tabular}                                                                                                 \\    \\
                                                                                                      &                                                                                                                             & Write the pseudocode and manually observe if the required output can be generated                                                                                                                                                                                                  \\         \\
                                                                                                      &                                                                                                                             & Implement a POC where necessary and if that suits your situation                                                                                                                                                                                                                            \\ \cmidrule(l){2-3} 
                                                                                                         & \multirow{3}{*}{\begin{tabular}[c]{@{}l@{}}Improve emotional intelligence \\ (self regulation)\end{tabular}}                 & \begin{tabular}[c]{@{}l@{}}Regulate low pleasurable emotions when human errors happen. \\ For example, consider mistakes as an opportunity to learn  \end{tabular}                                                                                                                  \\                                                                                                                            \\ \cmidrule(l){2-3} 
                                                                                                      & \multirow{2}{*}{\begin{tabular}[c]{@{}l@{}}Improve emotional \\ intelligence \\ (relationship management)\end{tabular}}     & Collaboratively work with other team members                                                                                                                                                                                                                                                \\ \\
                                                                                                      &                                                                                                                             & Be supportive to each other                                                                                                                                                                                                                                                                 \\ \cmidrule(l){2-3} 
                                                                                                      & \multirow{2}{*}{Improve agility}                                                                                            & Feel empowered/ autonomous/ responsible                                                                                                                                                                                                                                           \\          \\
                                                                                                      &                                                                                                                             & Minimise forcing the effort -- working overtime, as much as possible                                                                                                                                                                                                                           \\ \cmidrule(l){2-3} 
                                                                                                   
                                                                                                      & Improve cognitive intelligence                                                                                              & \begin{tabular}[c]{@{}l@{}}Be due diligent:\\ Ideate\\ Learn/ acquire new knowledge\end{tabular}                                                                                                 \\ \midrule
\multirow{3}{*}{Manager}                                                                              & \begin{tabular}[c]{@{}l@{}}Improve emotional intelligence \\ (social awareness)\end{tabular}                                & Feel the team and motivate them accordingly                                                                                                                                                                                                                                                 \\ \cmidrule(l){2-3} 
                                                                                                      & \multirow{3}{*}{\begin{tabular}[c]{@{}l@{}}Improve emotional intelligence \\ (relationship management)\end{tabular}}        & Promptly and honestly communicate with the team                                                                                                                                                                                                                                             \\ \\
                                                                                                      &                                                                                                                             & Have proper coordination and conversations with the team                                                                                                                                                                                                                                     \\ \midrule
\multirow{3}{*}{Customer}                                                                             & \begin{tabular}[c]{@{}l@{}}Improve emotional intelligence \\ (relationship management)\end{tabular}                         & Positively engage with the team                                                                                                                                                                                                                                                             \\ \cmidrule(l){2-3} 
                                                                                                      & \begin{tabular}[c]{@{}l@{}}Improve emotional intelligence \\ (social awareness and \\ relationship management)\end{tabular} & \begin{tabular}[c]{@{}l@{}}Appreciate the team by showing high pleasurable emotions when you are satisfied with \\ the work they delivered\end{tabular}                                                                                                                                     \\ \cmidrule(l){2-3} 
                                                                                                      & Know the business need                                                                                                      & If you are unsure about the RCs you request, talk with the team   \\ \bottomrule
\end{tabular}%
}
\end{table*}

\subsection{Implications for Researchers}






\label{imp:researchers}
\textbf{Rational Emotive Therapy: Antecedents, beliefs, and emotional consequences. } We have presented the factors that our study participants reported to heavily influence RC handling and in triggering emotions of the practitioners.  These findings were derived from the opinions of the practitioners -- which are what they believe but which may not be true. According to the ABC model of emotions, antecedents (factors) (A) lead to beliefs (B) that eventually result in consequences (including emotions) (C). This can take two forms: (1) antecedent $\rightarrow$ rational belief $\rightarrow$ rational emotions, or (2) antecedent $\rightarrow$ irrational belief $\rightarrow$ irrational emotions. 

Applying these two forms to our findings around low pleasurable emotions, it can be said that the low pleasurable emotions may not be always due to rational beliefs the practitioners have. For example, practitioners making mistakes generating low pleasurable emotions, can be thought of as an irrational belief; where mistakes could be considered as an opportunity to learn, thus resulting in high pleasurable emotions. Questioning the rationality around the beliefs. For instance, questioning the rationality around the beliefs that result in low pleasurable emotions, could help in identifying (1) the irrational beliefs and (2)  navigating the irrational belief to replace them with new beliefs that result in high pleasurable emotions. We encourage  researchers to investigate this further to devise potential strategies that could replace the irrational beliefs of practitioners -- i.e., how to regulate low pleasurable emotions (one of key aspects in emotional intelligence), so that preferred high pleasurable emotions are produced in them when handling RCs.

\textbf{Antecedent--focused emotion regulation.} While rational emotive therapy focuses on regulating the emotions by changing the beliefs, the emotions can be regulated using antecedent--focused emotion regulation (as opposed to response--focused emotion regulation). That is changing the antecedents that trigger the emotions. In simple terms, we are changing the situations early, so that emotions are regulated. For example, when the size of the RC is large (antecedent) and causes low pleasurable emotions, the RC can be broken down into manageable sizes (regulation) so that feeling low pleasurable emotions are regulated. The best practices we have given in Section \ref{imp:prac} are based on antecedent--focused emotion regulation. Further research on how to regulate emotions by managing the antecedents is encouraged.

\textbf{Emotions are not the only consequence of stakeholder factors.} In Section \ref{sec:stakeholder}, for some stakeholder factors, we identified other consequences than just emotions, though we have not found additional consequences for all the factors. We hypothesise that there could be other consequences for all factors and even additional consequences for the factors for which we found other consequences. We encourage researchers to investigate this in the future.

\textbf{Age, behavior, team climate, and emotional consequences.} Some of our findings indicate that participants perceived the age of the peers in the team impact the RC handling and result in the emotions of the practitioners. For example, $\beta$P156 said that old peers stick to traditional ways of working, resulting in low efficiency, thus making him feel low pleasurable emotions. On the other hand, some other practitioners mentioned that working with young peers bring high pleasurable emotions in them, and they find new ways to handle the work and do the work perfectly. 






A member of a team could have a variety of experiences and expertise related to work that is not only limited to RCs, but also to the other tasks carried out by the team. Therefore, it is discriminatory to form a team based on age, or consider that older peers are not suitable to have within the team but young peers are. If a particular individual is not flexible enough to change their ways of working or to gain new knowledge to adhere to the working mechanism in the team, it is the responsibility of the manager to help them to take the right measures. While Baltes et al. summarised most common employment strategies for older developers in their article \cite{Baltes2020IsDevelopers}, we have presented how behavior change models can be used in agile contexts to change the behavior of certain agile roles in a theory--based previous study of us \cite{Madampe2020TowardsModels}. However, we have not studied the practicality of the suggestions we have made. Therefore, we recommend the future researchers study this in the future, as  we consider this as a crucial area to address for the betterment of the software teams.

\textbf{Low pleasurable stakeholder factors as challenges.}
Even though the survey Beta participants stated how different stakeholders influence the RC handling and their emotions, it can be hypothesised that the factors that aroused low pleasurable emotions in them could be challenges to them. However, since they did not explicitly mention that they were challenges, we did not conclude them as challenges. This could be validated in the future to expand the challenge assessment metrics that we have given in this paper.

\section{Threats to Validity}

\textbf{External Validity.}
The majority of the participants in survey Alpha represented Asia, while the majority of the participants in survey Beta represented North America. The emotions felt for the challenges (answer to RQ2) reported by survey Alpha participants, were reported by the survey Beta participants. This has 2 aspects. (1) survey Alpha participants may have felt different to the challenges that they experienced, (2) survey Beta participants may have faced additional/less challenges than the ones that we identified in survey Alpha. However, as it was not possible to visit survey Alpha participants back query about the emotions felt, the first point mentioned above remains as a threat to the validity of the findings of RQ2.  As mentioned in Section \ref{imp:researchers}, the low pleasurable stakeholder factors could be considered as additional challenges faced by survey Beta practitioners. However, since this was not confirmed through our study, we did not derive any conclusions. Both surveys were conducted during the Covid-19 global pandemic, where the emotional wellbeing of every global citizen was impacted. The findings around the emotions we have presented in this paper may have been different during a time without a pandemic. If the global situation changes in the future, future researchers may consider replicating this study again to compare and contrast to validate the findings we have given in this paper.

\textbf{Construct Validity.}
In both surveys, we provided the definition of RC so that all participants share the same understanding. However, the definitions of the challenging factors in survey Alpha and the definitions for emotions in survey Beta were not given. We assumed that the participants share a common understanding of the terminology. However, this might not have been the case for every participant. Therefore, in future studies, we suggest researchers to include definitions as much as possible in the surveys.

\textbf{Reliability.}
All data were analysed by the first author. In order to mitigate the subjectivity of the analysis, the team had weekly meetings and discussed the findings. After iteratively going through the analysis, a final meeting was held between the first and the second author to finalise the findings. Hence, the researcher bias was mitigated.

\section{Conclusions}
Through findings from two survey studies, this paper presents the key challenging factors that make RC handling a challenge and how practitioners feel when RCs are challenging to handle, how practitioners themselves, their peers in the team, their managers, and customers influence the overall RC handling. Some challenges are technical and some are social which also belong to aspects of agile, emotional intelligence, and cognitive intelligence. Therefore, to better handle RCs with positive emotions in socio--technical environments, agility, emotional intelligence, and cognitive intelligence need to cooperate with each other.  

\section*{Acknowledgments}
This work is supported by a Monash Faculty of IT scholarship. Grundy is supported by ARC Laureate Fellowship FL190100035. Also, our sincere gratitude goes to Dr William Bingley for providing invaluable feedback for this work, and all the participants who took part in this study.

\bibliographystyle{ieeetr}
\bibliography{main}

\appendices
\onecolumn

\section{Project Information}

\begin{table*}[htp!]
\caption{Information of Current/Most Recent Project of the Survey Alpha and Beta Participants (Other: $\leq$ 5 participants)}
\label{tab:project}
\resizebox{\textwidth}{!}{%
\begin{tabular}{@{}llllll@{}}
\toprule
\multicolumn{6}{l}{\cellcolor[HTML]{EFEFEF}\textbf{Survey Alpha}}                                                              \\ \midrule

\textbf{Project Domain} & \textbf{\# of Participants}                        &  \textbf{Project   Category} &\textbf{\# of Participants}                                                         & \textbf{Agile Method Used} &\textbf{\# of Participants}                             \\ \midrule
IT                                                            & \mybarhhigh{2.3}{23} & New Development                                                                                 & \mybarhhigh{2.2}{22}    & Scrum                                                                 & \mybarhhigh{2.8}{28}   \\
Other                                                    & \mybarhhigh{1.7}{17}  & Software as a Service                                                                                            & \mybarhhigh{0.9}{9}     & Other                                                                & \mybarhhigh{1.2}{12}    \\
                                  &   & Other                                                                                       & \mybarhhigh{0.9}{9}     &                                                               &     \\
   \\ \cmidrule(l){1-4} 
                                                      \textbf{Team Size} & \textbf{\# of People}                                               & \textbf{Iteration Length} & \textbf{\# of Weeks}         &   &       \\ \cmidrule(l){1-4} 
                                                       Minimum                                                                                            & 3     & Minimum                                                                  & 2  &   &   \\
                                                 Maximum                                                                                            & 80    & Maximum                                                                  & 10   &   & \\
                                                            Mean                                                                                            & 14.36 & Mean                                                                  & 3.28   &    & \\
                                                              Standard Deviation                                                                                       & 15.54 & Standard Deviation                                                             & 2.25  &    & \\ \midrule
                                                              \multicolumn{6}{l}{\textbf{Practices Followed (Order of the Bars in Each Graph Below: Never $\rightarrow$ Sometimes $\rightarrow$ About half the time $\rightarrow$ Most of the Time $\rightarrow$ Always)}} \\ \midrule 
  Collective Estimation&\mybarX{0.03}{0.09}{0.06}{0.1}{0.12}&                          Product Backlog&\mybarX{0.01}{0.05}{0.06}{0.07}{0.21}&                               Scrum/Kanban Board&\mybarX{0.01}{0.03}{0.04}{0.06}{0.26}\\
  Customer Demos&\mybarX{0.01}{0.15}{0.05}{0.08}{0.11}&
  Short iterations/Sprints&\mybarX{0}{0.08}{0.03}{0.1}{0.19}&
  Self--assignment&\mybarX{0.02}{0.1}{0.06}{0.15}{0.07}\\
  Daily Standup/Team Meeting&\mybarX{0.01}{0.01}{0.02}{0.06}{0.3}&
  Release Planning&\mybarX{0}{0.05}{0.04}{0.11}{0.2}&
  Sprint Backlog&\mybarX{0.03}{0.09}{0.04}{0.07}{0.17}\\
  Definition of Done&\mybarX{0.02}{0.06}{0.04}{0.12}{0.16}&
  Retrospectives&\mybarX{0.03}{0.09}{0.05}{0.1}{0.13}&
User Stories&\mybarX{0.01}{0.02}{0.06}{0.12}{0.19}\\
 Iteration Planning&\mybarX{0.01}{0.06}{0.02}{0.14}{0.17}&
 Review Meetings&\mybarX{0}{0.09}{0.05}{0.09}{0.17}&
 Pair Programming&\mybarX{0.04}{0.17}{0.05}{0.1}{0.04}\\  \midrule
 
 \multicolumn{6}{l}{\cellcolor[HTML]{EFEFEF}\textbf{Survey Beta}}                                                              \\ \midrule
 \textbf{Project Domain} & \textbf{\# of Participants} & \textbf{Project Category} & \textbf{\# of Participants} & \textbf{Development Method Used} & \textbf{\# of Participants} \\ \midrule
IT                         & \mybarhhighY{12.2}{122}                         & New development           & \mybarhhighY{11.5}{115}                         & Scrum                            & \mybarhhighY{5.7}{57}                          \\
Finance \& Banking         & \mybarhhighY{3}{30}                          & Software as a Service     & \mybarhhighY{4.7}{47}                          & Dynamic System Development       & \mybarhhighY{4.4}{44}                          \\
Manufacturing              & \mybarhhighY{1}{10}                          & Maintenance               & \mybarhhighY{2.2}{22}                          & Feature Driven Development       & \mybarhhighY{2.5}{25}                          \\
Transport                  & \mybarhhighY{1}{10}                          & Migration                 & \mybarhhighY{1.7}{17}                          & Waterfall                        & \mybarhhighY{1.7}{17}                          \\
Telecom                    & \mybarhhighY{0.7}{7}                           &                           &                             & Kanban                           & \mybarhhighY{1.4}{14}                        \\
Other                   & \mybarhhighY{1.6}{16}                          &                           &                             & Crystal                          & \mybarhhighY{1.1}{11}                         \\ \cmidrule(r){1-4}
\textbf{Team Size }              & \textbf{\# of People}                & I\textbf{teration Length }         & \textbf{\# of Weeks}                 & None                             & \mybarhhighY{0.8}{8}                            \\ \cmidrule(r){1-4}
Minimum                 & 1                           & Minimum                   & 1                           & Other                            & \mybarhhighY{1.1}{11}                          \\
Maximum                 & 100                         & Maximum                   & 10                          &                                  &                             \\
Mean                    & 18.61                       & Mean                      & 5.67                        &                                  &                             \\
Standard Deviation      & 18.8                        & Standard Deviation        & 2.11                        &                                  &                                            \\ \midrule
\multicolumn{6}{l}{\textbf{Practices Followed   (Order of the Bars in Each Graph Below: Never $\rightarrow$ Sometimes $\rightarrow$ About half the time $\rightarrow$  Most of the time $\rightarrow$ Always)}}                            \\ \midrule
Collective Estimation      & \mybar2{0.3}{1.566666667}{1.366666667}{2.633333333}{0.833333333}               & Product Backlog           & \mybar2{0.333333333333333}{1.16666666666667}{1.33333333333333}{2.33333333333333}{1.53333333333333}              & Scrum/Kanban Board               & \mybar2{0.5}{1.03333333333333}{1.43333333333333}{1.96666666666667}{1.76666666666667}              \\
Customer Demos             & \mybar2{0.366666666666667}{1.23333333333333}{1.86666666666667}{2.1}{1.13333333333333}              & Short Iterations/Sprints  & \mybar2{0.333333333333333}{1.26666666666667}{1.56666666666667}{1.96666666666667}{1.56666666666667}              & Self-assignment                  & \mybar2{0.433333333333333}{1.16666666666667}{1.4}{2.3}{1.4}             \\
Daily Standup/team meeting & \mybar2{0.133333333333333}{0.933333333333333}{1.26666666666667}{1.86666666666667}{2.5}               & Release Planning          & \mybar2{0.2}{1.03333333333333}{1.43333333333333}{2.63333333333333}{1.4}               & Sprint Backlog                   & \mybar2{0.466666666666667}{1.5}{1.33333333333333}{2.06666666666667}{1.33333333333333}              \\
Definition of Done         & \mybar2{0.4}{1}{1.53333333333333}{2.33333333333333}{1.43333333333333}              & Retrospectives            & \mybar2{0.733333333333333}{1.4}{1.36666666666667}{1.83333333333333}{1.36666666666667}              & User Stories                     & \mybar2{0.533333333333333}{0.833333333333333}{1.56666666666667}{1.9}{1.86666666666667}              \\
Iteration Planning         & \mybar2{0.266666666666667}{0.833333333333333}{1.56666666666667}{2.6}{1.43333333333333}               & Review Meetings           & \mybar2{0.4}{1.06666666666667}{1.23333333333333}{1.93333333333333}{2.06666666666667}              & Use Cases                        & \mybar2{0.4}{0.8}{1.3}{2.53333333333333}{1.66666666666667}             \\
Pair Programming           & \mybar2{1.13333333333333}{1.5}{1.53333333333333}{1.6}{0.933333333333333}              &                           &                             &                                  &                             \\ \bottomrule
\end{tabular}%
}
\end{table*}

\clearpage

\section{Challenges and Emotions Felt -- Examples}

\begin{table*}[htp!]
\caption{Challenges and Emotions Felt (Dev: Developer; BA: Business Analyst; AC/SM: Agile Coach/ Scrum Master; Mgr: Manager; Sys. Admin: System Administrator) }
\label{tab:challenges_emotions}
\resizebox{0.885\textwidth}{!}{%
\begin{tabular}{@{}lllll@{}}
\toprule
\textbf{Challenge}                           & \textbf{Dimension}            & \textbf{Emotions}                                                                                                                   & \textbf{Examples}                                                                                                                                                                                                                                                                                                                                                                                                                                                                                                                                                                                                                                                                                                                                                                                                                                                                                                                                                                 & \textbf{Roles (\textless{}Participants\textgreater{})                                 }                                                                                                                      \\ \midrule
\multirow{2}{*}{Complexity}         & High                 & \begin{tabular}[c]{@{}l@{}}Low\textsuperscript{1} in gen.\\ Anxious\end{tabular}                                                            & \textit{\begin{tabular}[c]{@{}l@{}}``Client requested a new functionality that would need to make a \\ lot of changes in the project that was developed with a high level \\ of complexity'' -- $\beta$P82\\ \\ ``When we had to deal with additions and/or modifications that \\ were highly complex and/or unexpected, I tended to feel anxious'' \\ -- $\beta$P188\end{tabular}}  & Dev ($\beta$P82, $\beta$P128, $\beta$P188)   \\ \cmidrule(l){2-5} 
                                                                                    &                                   Low                  & \begin{tabular}[c]{@{}l@{}}High\textsuperscript{2} in gen.\\ Excited\\ Energetic\\ \\ High\textsuperscript{1} in gen.\\ Content\\ Calm\\ Relaxed\end{tabular} & \textit{\begin{tabular}[c]{@{}l@{}}``When the implementation turns out to be less complex than initially \\ specified'' -- $\beta$P94\\ \\ ``When the owner asked us change the third party service provider \\ which they insisted early on the project I felt energetic because we \\ could do it better and easier'' -- $\beta$P113\end{tabular}}                                                                                                                                                                                                                                                                                                                                                                                                                                                                                                                                                                                                                                & \begin{tabular}[c]{@{}l@{}}Dev ($\beta$P82, $\beta$P94, $\beta$P110, $\beta$P113, $\beta$P125, \\ $\beta$P128, $\beta$P188)\\ BA ($\beta$P170, $\beta$P183)\\ Tester ($\beta$P54)\\ AC/SM, Dev ($\beta$P76)\\ Dev, Mgr ($\beta$P116)\\ Dev, Sys. Admin ($\beta$P177)\end{tabular}                                                                                                                                                                                                                                                                                                                                                                                                                                                                                                                                                                                                                                                                           \\ \midrule
\multirow{2}{*}{Cascading impact}   & 

High                 & \begin{tabular}[c]{@{}l@{}}Low\textsuperscript{1} in gen.\\ Angry\\ \\ Discouraged\\ Fatigued\end{tabular}                                  & \textit{\begin{tabular}[c]{@{}l@{}}``Nightmarish changes that totally change the database design, \\ application design, program structure or involve a complete overhaul \\ of the application due to a basic assumption or understanding based \\ on which the application was developed being charged at the last \\ minute are extremely scary, time consuming, risky and difficult to \\ implement without  from scratch. These are changes that involve a \\ change in database structure, constraints and relationship between \\ files, program logic, etc. Usually, this will require a re-estimation, \\ and change in deadlines. Sometimes this even requires starting from \\ scratch to avoid design issues'' -- $\beta$P128\\ \\ ``When requirements were significantly changed in the middle of \\ the project, typically needing significant modification of work we'd \\ already done, I tended to feel discouraged \\ and fatigued'' -- $\beta$P188\end{tabular}} & Dev ($\beta$P128, $\beta$P138, $\beta$P151, $\beta$P188)  
                                                 \\ \cmidrule(l){2-5} 
                                    &                                  Low                  & \begin{tabular}[c]{@{}l@{}}High\textsuperscript{1} in gen.\\ Content\end{tabular}                                                           & \textit{\begin{tabular}[c]{@{}l@{}}``Some examples are the wordings on screens, minor layout changes, \\ additional variations, or changes that can be implemented without \\ major changes to the existing application \\ design or program flow.'' -- $\beta$P128\end{tabular}}                                                                                                                                                                                                                                                                                                                                                                                                                                                                                                                                                                                                                                                                                             & \begin{tabular}[c]{@{}l@{}}Dev ($\beta$P128, $\beta$P130) \\ BA, Dev, Tester ($\beta$P92)\end{tabular}                                                                                                                                                  \\ \midrule
Size                                & Large                & Low\textsuperscript{2} in gen.                                                                                                              & \textit{\begin{tabular}[c]{@{}l@{}}``Large new features being added, that necessitate large rewrites \\ as they were well beyond the initial scope'' -- $\beta$P10\end{tabular}}                                                                                                                                                                                                                                                                                                                                                                                                                                                                                                                                                                                                                                                                                                                                                                                              & Dev ($\beta$P10)                                                                                                                                                                                          \\ \midrule
\multirow{2}{*}{Definition}         & Precise/ clear       & High\textsuperscript{1} in gen.                                                                                                             & \textit{\begin{tabular}[c]{@{}l@{}}``well--defined feature definitions'' -- $\beta$P174\\ \\ ``When we finally got UI mockups, I was pleased that I knew exactly \\ what I was supposed to be building, instead of having to design the \\ UX as I went along. I was also happy all the details (fonts, spacing, \\ etc.) were exactly specified, so I could code it right the first time \\ instead of having to do a second pass for consistency with the rest \\ of the UI'' -- $\beta$P175\end{tabular}}                                                                                                                                                                                                                                                                                                                                                                                                                                                                        & \begin{tabular}[c]{@{}l@{}}Dev ($\beta$P175)\\ Dev, Mgr ($\beta$P174)\end{tabular}                                                                                                                              \\ \cmidrule(l){2-5} 
                                    & Imprecise/ unclear   & Low\textsuperscript{2} in gen.                                                                                                              & \textit{\begin{tabular}[c]{@{}l@{}}``I just hate projects where I need to run after information cause \\ nobody knows anything, it tends to put me in a negative mood'' \\ -- $\beta$P104\end{tabular}}                                                                                                                                                                                                                                                                                                                                                                                                                                                                                                                                                                                                                                                                                                                                                                       & Dev ($\beta$P104)                                                                                                                                                                                         \\ \midrule
\multirow{2}{*}{Access to customer} & Easy/ regular        & Calm                                                                                                                      & \textit{\begin{tabular}[c]{@{}l@{}}``Requirements for permitting issuance manual refunds changed \\ to require a not more fraud detection code. I was calm when we had \\ this change, because we have regular conversations with the \\ stakeholder who the project was for. With the regular collaboration, \\ I knew that she would understand delay at not be angry at us for \\ this'' -- $\beta$P184\\ \\ ``satisfied. Customer input happens throughout the development \\ process'' -- $\beta$P187\end{tabular}}                                                                                                                                                                                                                                                                                                                                                                                                                                                            & \begin{tabular}[c]{@{}l@{}}Dev, Mgr ($\beta$P187)\\ Dev, Tester ($\beta$P184)\\ AC/SM, Dev, Tester ($\beta$P83)\end{tabular}                                                                                          \\ \cmidrule(l){2-5} 
                                    & Difficult/ irregular & \begin{tabular}[c]{@{}l@{}}Fatigued\\ Gloomy\end{tabular}                                                                 & \textit{\begin{tabular}[c]{@{}l@{}}``well we used Dynamic systems development method in our \\ project as the client needed an updation on their application as \\ the are shifting to an newer location we used Rapid Application \\ Development using DSDM. Overall it was an fatigued experience \\ cause project seems to be moving a bit gloomy as client was \\ always busy so reply was much delayed.'' -- $\beta$P165\\ \\ ``Late in the dev phase of a new work project, my client changed \\ their requirements; specifically they wanted to change how their \\ data is loaded into the application, changing it from an XLS file \\ to a database table.. I had to rebuild the file loader completely. \\ It made me feel angry because I felt they do not understand my \\ position, did not take our meetings seriously or read my \\ emails clearly to prevent this from happening'' -- $\beta$P150\end{tabular}}                                                    & \begin{tabular}[c]{@{}l@{}}Dev ($\beta$P125, $\beta$P165)\\ Dev, Tester ($\beta$P150)\end{tabular}                                                                                                                    \\ \midrule
Effort required                     & High                 & Low\textsuperscript{2} in gen.                                                                                                              & \textit{\begin{tabular}[c]{@{}l@{}}``The client asked for an unexpected change when we had finished \\ loading more than 25gbs of data and also all the databases were \\ already connected. We had to make a change, call the engineers \\ and rewrite more than 256 lines of code and upload everything \\ back to the two servers. The change was basically that the he \\ wanted the  app to be hybrid. It was an app for mobile payments. \\ The client wanted the app to work with all devices, when at the \\ beginning he had asked for it only for android. The conversion was \\ not difficult, the problem was to delete all files from the server and \\ upload everything again. It was a headache'' -- $\beta$P72\\ \\ ``These emotions were experienced when requirements are changed \\ that will increase the amount of effort required significantly'' -- $\beta$P123\end{tabular}}                                                                               & \begin{tabular}[c]{@{}l@{}}Dev ($\beta$P123)\\ AC/SM, Dev ($\beta$P72)\end{tabular}                                                                                                                             \\ \bottomrule
\end{tabular}%
}
\end{table*}

\clearpage

\section{Practitioner--related Factors -- Examples}

\begin{table*}[ht!]
\caption{Practitioner--related factors (IC1--IC4: Further Consequences; in gen.: in general; Dev: Developer; Mgr: Manager; AC/SM: Agile Coach/Scrum Master; BA: Business Analyst; PO: Product Owner)}
\label{tab:practitioner}
\resizebox{\textwidth}{!}{%
\begin{tabular}{@{}llll@{}}
\toprule
\textbf{\faBrain Factor}                                                                                                          & \textbf{Emotions}                                                                                           & \textbf{Examples}                                                                                                                                                                                                                                                                                                                                                                                                                                                                                                                                                                                                                                       & \textbf{Roles (\textless{}Participants\textgreater{})}                                                                                                                                                             \\ \midrule
\begin{tabular}[c]{@{}l@{}}{[}+A{]} Preference in learning/ \\ knowledge acquisition \textsuperscript{{[}IC2, IC3, IC4{]}}\end{tabular}     & \begin{tabular}[c]{@{}l@{}}High\textsuperscript{2} in gen.\\ Excited\\ Inspired\end{tabular}                                  & \textit{\begin{tabular}[c]{@{}l@{}}``Because I was assigned with the changes related to the component I was \\ responsible, and it was interesting because I had to learn new concepts'' \\ -- $\beta$P25\\ \\ ``I was working on project with dd payment method and they included the \\ concept of virtual payment so I had to do the webservice and linking the \\ payment bank account with our own webservice to capture the data which \\ was fun and excitement for me do it. It was new and I learned a lot out of \\ it which is great'' -- $\beta$P126\end{tabular}}                                                                                               & \begin{tabular}[c]{@{}l@{}}Dev ($\beta$P106, $\beta$P199)\\ Dev, Tester ($\beta$P25, \\ $\beta$P88, $\beta$P134)\\ BA ($\beta$P189)\\ BA, Dev ($\beta$P126)\\ BA, Dev, Mgr ($\beta$P143)\\ AC/SM, Dev ($\beta$P103)\end{tabular} \\ \midrule
\begin{tabular}[c]{@{}l@{}}{[}+B{]} High empowerment/ auntonomy/ \\ responsibility \textsuperscript{{[}IC1{]}}\end{tabular}                 & \begin{tabular}[c]{@{}l@{}}Energetic\\ Inspired\\ \\ High\textsuperscript{1} in gen.\\ Satisfied\\ Relaxed\end{tabular}       & \textit{\begin{tabular}[c]{@{}l@{}}``The combination of autonomy, pay, flexibility, and job satisfaction leads \\ to software engineers being quite happy compared to people in other \\ professions'' -- $\beta$P195\\ \\ ``I was placed as a head tech on a major project to save our company \\ thousands of dollars. Knowing that I could potentially save our company \\ over 1 million dollars with time saving and energy saving techniques \\ made me quite excited. I was really energetic and felt inspired to get the \\ best information I could to make sure our company could do the same \\ amount of work but use less energy'' -- $\beta$P196\end{tabular}} & \begin{tabular}[c]{@{}l@{}}BA ($\beta$P195)\\ Mgr ($\beta$P107)\\ AC/SM, Tester \\ ($\beta$P196)\end{tabular}                                                                                \\ \midrule
{[}+C{]} Practitioner ideating                                                                                            & High\textsuperscript{2} in gen.                                                                                               & \textit{\begin{tabular}[c]{@{}l@{}}``I felt these emotions with the ideas that came to my mind, about the \\ new project I am developing'' -- $\beta$P81\\ \\ ``When an addition to the functionality the costumer wants is something \\ you thought about before as some kind of exercise'' -- $\beta$P94\end{tabular}      }                                                                                                                                                                                                                                                                                                                                               & \begin{tabular}[c]{@{}l@{}}Dev ($\beta$P94)\\ PO ($\beta$P81)\\ AC/SM, Mgr, Tester \\ ($\beta$P133)\end{tabular}                                                                             \\ \midrule
\begin{tabular}[c]{@{}l@{}}{[}-A{]} Difficulty in learning/ knowledge \\ acquisition\end{tabular}                         & \begin{tabular}[c]{@{}l@{}}Anxious\\ \\ Gloomy\\ Fatigued\\ Discouraged\\ Depressed {[}over{]}\end{tabular} & \textit{\begin{tabular}[c]{@{}l@{}}``Anxious when i feel that the new technology is difficult to learn'' \\ -- $\beta$P103\\ \\ ``During the initial learning of clients modification i felt very anxious \\ during the learning procedure of {[}project management tool{]} + \\ Agile on the implementation'' -- $\beta$P67\end{tabular} }                                                                                                                                                                                                                                                                                                                                  & \begin{tabular}[c]{@{}l@{}}Mgr ($\beta$P66)\\ AC/SM ($\beta$P67)\\ AC/SM, Dev \\ ($\beta$P103)\\ Tester, Product \\ Specialist ($\beta$P135)\end{tabular}                                          \\ \midrule
\begin{tabular}[c]{@{}l@{}}{[}-B{]} Less autonomy due to forced effort \\ on working overtime \textsuperscript{{[}IC5, IC6{]}}\end{tabular} & \begin{tabular}[c]{@{}l@{}}Low\textsuperscript{1} in gen.\\ \\ Low\textsuperscript{2} in gen.\\ Bored\\ Depressed\end{tabular}                  & \textit{\begin{tabular}[c]{@{}l@{}}``overwork , get stuck due to dependency on other developments, \\ only solve bugs, not see progress in the project ... redo the same \\ story many times'' -- $\beta$P106\\ \\ ``Depression on the job is often misinterpreted as a bad attitude or \\ poor work ethic I feel sometimes depressed because of over time \\ work and excess of project on my head, due date is to near to me, \\ so that I lost my sleep and my relax time. so i felt to much depressed'' \\ -- $\beta$P147\end{tabular} }                                                                                                                                 & \begin{tabular}[c]{@{}l@{}}Dev ($\beta$P56, $\beta$P86, \\ $\beta$P89, $\beta$P106)\\ Dev, Tester \\ ($\beta$P192)\\ BA ($\beta$P132, \\ $\beta$P147)\\ BA, Dev ($\beta$P126)\\ AC/SM ($\beta$P129)\end{tabular}                 \\ \midrule
{[}-D{]} Making mistakes                                                                                                  & \begin{tabular}[c]{@{}l@{}}Low\textsuperscript{1} in gen.\\ Depressed {[}over{]}\end{tabular}                                 & \textit{\begin{tabular}[c]{@{}l@{}}``whenever i missed just liltle mistake like forget to add a semicolon \\ lol'' -- $\beta$P133\\ \\ ``some time over depression to run the program and finding the new \\ code for gaining our knowledge and developing the advanced \\ technology and identify the mistake easy replaced'' -- $\beta$P65\end{tabular} }                                                                                                                                                                                                                                                                                                                  & \begin{tabular}[c]{@{}l@{}}Dev ($\beta$P131)\\ Tester ($\beta$P65)\\ BA, PO, Tester \\ ($\beta$P141)\\ Mgr, PO, Tester \\ ($\beta$P155)\end{tabular}                                               \\ \bottomrule
\end{tabular}%
}
\end{table*}

\clearpage

\section{Peer--related Factors -- Examples}

\begin{table*}[htp!]
\caption{Peer--related factors (PC1--PC5: Additional Consequence of the Antecedent; in gen.: in general; Dev: Developer; Mgr: Manager; AC/SM: Agile Coach/Scrum Master; BA: Business Analyst; PO: Product Owner)}
\label{tab:team}
\resizebox{\textwidth}{!}{%
\begin{tabular}{@{}llll@{}}
\toprule
\textbf{\faUsers Factor}                                                                                                                                       & \textbf{Emotions}                                                                                                                & \textbf{Examples}                                                                                                                                                                                                                                                                                                                                                                                                                                                                                                                                                                                                                                                                                                                                                                                                                                                          & \textbf{Roles (\textless{}Participants\textgreater{})}                                                                                                                                                             \\ \midrule
\begin{tabular}[c]{@{}l@{}}{[}+A{]} Having the required expertise/ competency/ \\ relevant experience of the peers\end{tabular}                        & \begin{tabular}[c]{@{}l@{}}Inspired\\ \\ High\textsuperscript{1} in gen.\\ Relaxed\\ Satisfied\\ Calm\\ At ease\end{tabular}                       & \textit{\begin{tabular}[c]{@{}l@{}}``I feel inspired in my project work team members. It help of my \\ team members and my project team member is very \\ knowledgeable persons'' -- $\beta$P107\\ \\ ``.. I belonged to an experienced team. Updating and changing \\ requirements - just as with any project - is something we had \\ done several times before this particular project. We (mostly) \\ knew what to expect and how to handle any unscheduled \\ issues or errors. We had every reason to stay calm and implement \\ changes in the same way we had done multiple times before'' \\ -- $\beta$P101\end{tabular}  }                                                                                                                                                                                                                                                            & \begin{tabular}[c]{@{}l@{}}Dev ($\beta$P61)\\ Tester ($\beta$P54)\\ Mgr ($\beta$P101, $\beta$P107)\\ Dev, Mgr ($\beta$P174)\end{tabular}                                                                 \\ \midrule
{[}+B{]} High efficiency/ productivity of peers                                                                                                        & \begin{tabular}[c]{@{}l@{}}High\textsuperscript{2} in gen.\\ Inspired\\ Excited\\ Energetic\\ \\ Relaxed\\ Satisfied\\ Calm\\ At ease\end{tabular} & \textit{\begin{tabular}[c]{@{}l@{}}``its combination of doing work that is doing work efficient the \\ output come also very well  once we do energetic work we get \\ excited result that result give enthusiastic state these all are come \\ we inspired some thing on our work that inspire we find'' -- $\beta$P176\\ \\ ``At sometime, I feel more at-ease whenever there is a change in \\ requirements from our client side because of my team members \\ as they are more efficient in their work and sometimes they don't \\ need any presence of mine in that time which makes me feel \\ at-ease'' -- $\beta$P109\end{tabular}   }                                                                                                                                                                                                                                                  & \begin{tabular}[c]{@{}l@{}}Dev ($\beta$P176, $\beta$P156)\\ Mgr ($\beta$P186, $\beta$P167, $\beta$P107)\\ Dev, Mgr ($\beta$P193)\end{tabular}                                                                  \\ \midrule
\begin{tabular}[c]{@{}l@{}}{[}+C{]} Better relationship management within the \\ team through collaboration and engagement \textsuperscript{{[}PC1--PC5{]}}\end{tabular} & \begin{tabular}[c]{@{}l@{}}High\textsuperscript{2} in gen.\\ Inspired\\ Energetic\\ \\ High\textsuperscript{1} in gen.\\ Relaxed\\ Satisfied\\ Calm\end{tabular}     & \textit{\begin{tabular}[c]{@{}l@{}}``When the team is functioning well, management is \\ communicating, and the revision process is flowing smoothly, I \\ feel relaxed and satisfied that we're going to meet objectives \\ on time and on budget'' -- $\beta$P198\\ \\ ``As we have the MVVM framework, the change of UI is \\ concentrated on a small number of source code files. The \\ changes are still within the scope of my work. Furthermore, \\ the requirement changes follow standard SOP and Scrum \\ process. We had a meeting to break down the requirements \\ into tasks, prioritise user stories and commit tasks within our \\ working hours. The changes also went through a proper \\ discussion between all team members, managers and \\ software architect to estimate the effort and the impact of \\ the changes on production systems'' -- $\beta$P3\end{tabular}} & \begin{tabular}[c]{@{}l@{}}Dev ($\beta$P3, $\beta$P96, $\beta$P118, $\beta$P160, \\ $\beta$P198)\\ BA ($\beta$P117)\\ Mgr ($\beta$P37, $\beta$P68, $\beta$P107)\\ Dev, Mgr ($\beta$P38, $\beta$P41) \\ Dev, Mgr, Tester ($\beta$P100)\end{tabular} \\ \midrule
{[}-B{]} Low efficiency/ low productivity of the team                                                                                                  & \begin{tabular}[c]{@{}l@{}}Low\textsuperscript{1} in gen.\\ Angry\\ Anxious\end{tabular}                                                           & \textit{\begin{tabular}[c]{@{}l@{}}``slow co-workers'' -- $\beta$P133\\ \\ ``I feel angry anxious anytime I work with old folks who \\ always have a slow work rate and always believes in their \\ traditional way of performing duties I am always furious and \\ always hope I get younger teammates'' -- $\beta$P156\end{tabular}                                                                                         }                                                                                                                                                                                                                                                                                                                                                                                                                                                                 & \begin{tabular}[c]{@{}l@{}}Dev ($\beta$P110, $\beta$P156)\\ AC/SM, Mgr, Tester ($\beta$P133)\end{tabular}                                                                                    \\ \midrule
{[}-D{]} Team members making mistakes                                                                                                                  & \begin{tabular}[c]{@{}l@{}}Low\textsuperscript{1} in gen.\\ Angry\\ Disgusted\end{tabular}                                                         & \textit{\begin{tabular}[c]{@{}l@{}}``when something developed not working because of someone \\ else did a mistake'' -- $\beta$P53\\ \\ Sometimes teammate make mistakes on very important projects. \\ These t$\beta$Pes of emotions are faced {[}by{]} all guys'' -- $\beta$P186\end{tabular}                                                                                         }                                                                                                                                                                                                                                                                                                                                                                                                                                                                                                             & \begin{tabular}[c]{@{}l@{}}Dev ($\beta$P53)\\ Mgr ($\beta$P186)\\ AC/SM ($\beta$P197)\end{tabular}                                                                                           \\ \bottomrule
\end{tabular}%
}
\end{table*}

\clearpage

\section{Manager--related Factors -- Examples}

\begin{table*}[htp!]
\caption{Manager--related factors (in gen.: in general; Dev: Developer; Mgr: Manager; AC/SM: Agile Coach/Scrum Master; BA: Business Analyst; PO: Product Owner)}
\label{tab:my-manager}
\resizebox{\textwidth}{!}{%
\begin{tabular}{@{}llll@{}}
\toprule
\textbf{\faUserSecret Factor}                                                                                                                                              & \textbf{Emotions}                                                                     & \textbf{Examples}                                                                                                                                                                                                                                                                                                                                                                                                                                                                                                            & \textbf{Roles (\textless{}Participants\textgreater{})}                                                                            \\ \midrule
{[}+A{]} Manager having empathy (social awareness)                                                                                                            & \begin{tabular}[c]{@{}l@{}}Energetic\\ \\ High\textsuperscript{1} in gen.\\ Calm\end{tabular}              & \textit{\begin{tabular}[c]{@{}l@{}}``Because I was never given a pressure on completion due date and I \\ was never given a negative reviews on the ongoing project day by \\ day'' -- $\beta$P132\\ \\ ``one spotify project our team member is mistake, we find the mistake \\ last time , so immediate work day night  we finish the project. and very \\ calm, because our new  team leader says  every project {[}is{]} a lesson'' \\ -- $\beta$P45\end{tabular}}                                                                   & \begin{tabular}[c]{@{}l@{}}BA ($\beta$P132)\\ BA, Dev ($\beta$P200)\\ Mgr ($\beta$P45)\end{tabular}         \\ \midrule
\begin{tabular}[c]{@{}l@{}}{[}+B{]} Manager building better relationships through \\ communication and coordination (relationship \\ management)\end{tabular} & \begin{tabular}[c]{@{}l@{}}Inspired\\ \\ Satisfied\\ Relaxed\end{tabular} & \textit{\begin{tabular}[c]{@{}l@{}}``My manger makes me feel inspired whenever he speaks with me'' \\ -- $\beta$P61\\ \\``great project managers are great communicators with high emotional \\ intelligence. prompt and honest communication immediately gives a \\ project manager satisfied and relaxed'' -- $\beta$P153 \end{tabular}}                                                                                                                                                                                               & \begin{tabular}[c]{@{}l@{}}Dev ($\beta$P61)\\ BA, Dev, Mgr ($\beta$P143)\\ AC/SM ($\beta$P153)\end{tabular} \\ \midrule
{[}-A{]} Manager lacking empathy (social awareness)                                                                                                           & \begin{tabular}[c]{@{}l@{}}Low\textsuperscript{1} in gen.\\ Angry\\ \\ Low\textsuperscript{2} in gen.\end{tabular}        & \textit{\begin{tabular}[c]{@{}l@{}}``my ex team leader was suddenly angry  for every minor mistake'' \\ -- $\beta$P45\\ \\ ``I worked more and more time each day to try to succesfully make \\ the job, and new issues appeared every day that needed more and \\ more time. We were all at same feelings but team leader didn't hear \\ his team feeling. He requested us to perform this project on at \\ 'just in time'. I lost energy and I worked whithout any positive \\ feedback. I lost my flow'' -- $\beta$P173\end{tabular}} & \begin{tabular}[c]{@{}l@{}}Dev ($\beta$P173)\\ BA ($\beta$P147)\\ Mgr ($\beta$P45)\end{tabular}             \\ \bottomrule
\end{tabular}%
}
\end{table*}

\clearpage

\section{Customer--related Factors -- Examples}

\begin{table*}[hpt!]
\caption{Customer--related Factors (CC1--CC3: Additional Consequence of the Antecedent; in gen.: in general; Dev: Developer; Mgr: Manager; AC/SM: Agile Coach/Scrum Master; BA: Business Analyst; PO: Product Owner)}
\label{tab:customer}
\resizebox{\textwidth}{!}{%
\begin{tabular}{@{}llll@{}}
\toprule
\textbf{\faPeopleArrows Factor}                                                                                                                            & \textbf{Emotions}                                                                                                  & \textbf{Examples}                                                                                                                                                                                                                                                                                                                                                                                                                                                                & \textbf{Roles (\textless{}Participants\textgreater{})}                                                                                                                                                        \\ \midrule
\begin{tabular}[c]{@{}l@{}} [+A] Customers' manifested high pleasurable \\ emotions {[}satisfied/ excited{]} \textsuperscript{{[}CC1{]}}\end{tabular}      & \begin{tabular}[c]{@{}l@{}}High\textsuperscript{2} in gen.\\ Energetic\\ Inspired\\ \\ High\textsuperscript{1} in gen.\\ Calm\\ Satisfied\end{tabular} & \textit{\begin{tabular}[c]{@{}l@{}}``I normally have these emotions when my clients love my work'' \\ -- $\beta$P116\\ \\ ``During  the software development, there was  demo for it. At the \\time we wanted to show the project demo differently with Visme \\and Covo tool. Everyone appreciated that it was so special. \\We felt so much energy, inspiration at the time'' -- $\beta$P158\end{tabular} }                                                                                         & \begin{tabular}[c]{@{}l@{}}Dev ($\beta$P98, $\beta$P110)\\ Dev, Mgr ($\beta$P116)\\ AC/SM, Dev ($\beta$P158)\end{tabular}                                               \\ \midrule
\begin{tabular}[c]{@{}l@{}}{[}+B{]} Customers' positive engagement with \\ the team\end{tabular}                                            & \begin{tabular}[c]{@{}l@{}}High\textsuperscript{2} in gen.\\ \\ Calm\\ Satisfied\end{tabular}                                        & \textit{\begin{tabular}[c]{@{}l@{}}``When the client understood a necessary change and we were able \\ to work together to define the requirement change. It would lead \\ to a better outcome'' -- $\beta$P12\\ \\ ``Customer input happens throughout the development process'' \\ -- $\beta$P187\end{tabular} }                                                                                                                                                                                    & \begin{tabular}[c]{@{}l@{}}BA, Dev, Mgr, \\ PO, Tester ($\beta$P12)\\ AC/SM, Dev ($\beta$P158)\\ Dev, Mgr ($\beta$P187)\end{tabular} \\ \midrule
\begin{tabular}[c]{@{}l@{}}{[}-A{]} Customers' manifested low pleasurable \\ emotions {[}unsatisfied/ unimpressed{]} \textsuperscript{{[}CC2{]}}\end{tabular} & \begin{tabular}[c]{@{}l@{}}Angry\\ \\ Low\textsuperscript{2} in gen.\\ Depressed\end{tabular}                                        & \textit{\begin{tabular}[c]{@{}l@{}}``These emotions come when you see the client is not satisfied \\ with the work you've presented to him /her'' -- P116\\ \\ ``Suppose our project {[}did{]} not impress client, then we feel very \\ bad and depressed'' -- P186\end{tabular}       }                                                                                                                                                                                                  & \begin{tabular}[c]{@{}l@{}}Dev, Mgr ($\beta$P116)\\ BA, Dev ($\beta$P126)\\ Mgr ($\beta$P186)\end{tabular}                                            \\ \midrule
{[}-C{]} Customer being sceptical \textsuperscript{{[}CC2, CC3{]}}                                                                                            & \begin{tabular}[c]{@{}l@{}}Low\textsuperscript{1} in gen.\\ Anxious\\ \\ Low\textsuperscript{2} in gen.\\ Bored\\ Depressed\\ Discouraged\end{tabular} & \textit{\begin{tabular}[c]{@{}l@{}}``If client is not clear about what he wants and ask for same \\ functional or non-fuctional changes again and again that makes \\ me discouraged and bored'' -- $\beta$P109\\ \\ ``Multiple changes on the same functionnality/aspect, clearly \\ showing the client does not know what {[}client{]} wants. You do this \\ change fully expecting that the developpement won't 'stick' and \\ you'll have to change this again soon'' -- $\beta$P77\end{tabular}} & \begin{tabular}[c]{@{}l@{}}Dev ($\beta$P77)\\ BA ($\beta$P90, $\beta$P109)\\ Tester ($\beta$P180)\\ Dev ($\beta$P199)\end{tabular}                                        \\ \bottomrule
\end{tabular}%
}
\end{table*}

\end{document}